\documentclass[aps,twocolumn,superscriptaddress,longbibliography,nofootinbib,prc]{revtex4-1}
\usepackage[colorlinks, urlcolor=blue,linkcolor=blue, anchorcolor=blue, citecolor=blue]{hyperref}
\usepackage{amssymb}
\usepackage{amsmath}
\usepackage{graphicx}
\usepackage{nicefrac}
\usepackage{slashed} 
\usepackage{dsfont}
\usepackage[english]{babel}
\usepackage[utf8]{inputenc}
\usepackage[T1]{fontenc}

\newcommand{\Nc}{N_{\rm c}}
\newcommand{\km}{k_-}
\newcommand{\kp}{k_+}
\newcommand{\lnf}{l_{\rm 1f}}
\newcommand{\lif}{l_{\rm 2f}}
\newcommand{\ltf}{l_{\rm 3f}}
\newcommand{\sign}{\mathop{\mbox{sign}}}
\newcommand{\im}{\mathop{\mbox{Im}}}
\newcommand{\re}{\mathop{\mbox{Re}}}

\newcommand{\CF}{C_{_{\rm F}}}
\newcommand{\Nf}{n_{\!f}}
\newcommand{\muB}{\mu_{\rm B}}

\def\bm#1{\text{\boldmath$#1$}}
\renewcommand{\vec}{\bm}
\renewcommand{\log}{\ln}

\def\pd{\partial} 
\def\dd{{\rm d}}
\def\sNN{s_{_{\rm NN}}}
\def\nF{f_{_{\rm F}}}
\def\nB{f_{_{\rm B}}}
\def\Im{\hbox{Im}}

\def\eg{e.\,g.\ }	
\def\ie{i.\,e.\ }	

\def\l{\left}			\def\r{\right}

\def\be{\begin{eqnarray}}       \def\ee{\end{eqnarray}}
\def\bea{\begin{eqnarray}}      \def\eea{\end{eqnarray}}
\def\bean{\begin{eqnarray*}}    \def\eean{\end{eqnarray*}}
\def\eq#1{(\ref{#1})}        \def\nonu{\nonumber}

\newcommand{\sumint}[1]{{\hbox{$\textstyle\sum$}\!\!\!\!\!\!\!\int\,}_{\!\!\!\!\raise-0.5ex\hbox{$\scriptstyle{#1}$}}}

\def\gsim{\mathrel{\rlap{\lower0.25em\hbox{$\sim$}}\raise0.2em\hbox{$>$}}} 
\def\lsim{\mathrel{\rlap{\lower0.25em\hbox{$\sim$}}\raise0.2em\hbox{$<$}}}
\def\lg{\mathrel{\rlap{\lower0.25em\hbox{$>$}}\raise0.25em\hbox{$<$}}}
\def\gl{\mathrel{\rlap{\lower0.25em\hbox{$<$}}\raise0.25em\hbox{$>$}}}

\graphicspath{{figs}}

\begin{document}

\title{Dilepton production at next-to-leading order and intermediate invariant-mass observables}

\author{Jessica Churchill}
 \affiliation{%
   Department of Physics, McGill University, 
   3600 University Street, Montreal, QC, Canada H3A 2T8}
\author{Lipei Du}
\affiliation{%
    Department of Physics, McGill University, 
    3600 University Street, Montreal, QC, Canada H3A 2T8}
\author{Charles Gale}
 \affiliation{%
   Department of Physics, McGill University, 
   3600 University Street, Montreal, QC, Canada H3A 2T8}
\author{Greg Jackson}
 \affiliation{%
   Institute for Nuclear Theory, Box 351550, 
   University of Washington, Seattle, WA 98195-1550, United States}
\affiliation{%
   SUBATECH, 
   Nantes Universit\'e, IMT Atlantique, IN2P3/CNRS,
4 rue Alfred Kastler, La Chantrerie BP 20722, 44307 Nantes, France}
\author{Sangyong Jeon}
 \affiliation{%
   Department of Physics, McGill University, 
   3600 University Street, Montreal, QC, Canada H3A 2T8}

\date{\today}

\begin{abstract}
The thermal QCD dilepton production rate is calculated at next-to-leading order in the strong coupling and at finite baryon chemical potential. The two-loop virtual photon self-energy is evaluated using finite temperature field theory and combined consistently with the self-energy in the Landau-Pomeranchuk-Migdal regime. We present new results for a dense baryonic plasma. The rates are then integrated using (3+1)-dimensional fluid-dynamical simulations calibrated to reproduce hadronic experimental results obtained at RHIC at energies ranging from those of the Beam Energy Scan to $\sqrt{s_{_{\rm NN}}} = 200$ GeV. We elaborate on the ability of dileptons to relay information about the plasma baryonic content and temperature.
\end{abstract}

\maketitle

\section{Introduction}

Heavy-ion collisions make it possible to study the strong nuclear force at 
temperatures and densities where the quark-gluon plasma (QGP) can exist \cite{[{See, for example, }][{, and references therein.}]Harris:2023tti}. 
Photons produced within the plasma are unobscured by the rest of the medium,
making them (conceptually) ideal probes~\cite{Shuryak:1978ij,Pisarski1981,Kajantie:1981wg,Hwa:1985xg,Kajantie:1986dh}.
Those include both real and virtual photons; the latter of which 
decay into on-shell lepton-antilepton pairs. 
The electromagnetic radiation can originate from {\em all} stages of the nucleus-nucleus interaction, 
\eg\ reactions involving initial (hard) partons would populate the high energy part of the spectrum
and hadron decays (mostly $\pi^0$'s) would dominate the lower energy regions. 
Thus, a detailed understanding of specific production rates---particularly from the QGP itself---is 
a prerequisite to explaining the observations in general \cite{[{See, for example, }][{, and references therein.}]Gale:2009gc}, and to turning electromagnetic radiation into useful calibrated probes of the strongly interacting medium. 

Although dileptons are typically rarer than photons---their emission is suppressed by one power of $\alpha_{\rm em}$ compared to that of real photons---they have an additional degree of freedom:
the (non-zero) invariant mass, $M$.
This extra parameter is important in enabling 
dileptons as a tomographic tool of the medium created in heavy ion collisions. In this capacity, lepton pairs are a good probe of the local temperature, as the differential lepton pair emission rate depends on the invariant mass, $M$, which is independent of the local fluid flow~\cite{Rapp:2014hha}. 
This is not the case for real photons\footnote{See, however, the recent study of Ref. \cite{Paquet:2023bdx}.}~\cite{vanHees:2011vb}, but that dependence on the fluid flow can be exploited to inform the dynamical modeling of the collision~\cite{Shen:2013vja}. In the context of heavy-ion phenomenology, measurements of real and virtual photons are therefore complementary. 

Concerning thermal dilepton production,\footnote{%
Of course, the generation of electromagnetic probes from  early pre-equilibrium sources also deserves careful attention~\cite{Strickland1994,Srivastava1998,Greif:2016jeb,Hauksson:2017udm,Churchill2020,Gale:2021emg,Coquet:2021lca}.}
as the system evolves,
 QGP sources will compete with those associated with reactions involving composite hadrons such as mesons and baryons~\cite{Gale1987,Gale1993}. The partonic reaction rates---relevant at high temperature---and the hadronic reaction rates---important at lower temperature---are integrated using approaches simulating the time evolution of the entire strongly interacting system.  
Some of the early phenomenological studies were based on 
parton cascades~\cite{Geiger1992} and 
quasiparticle models~\cite{AP}, on a background of  
ideal hydrodynamics with radial symmetry~\cite{Huovinen2002}, and on thermal fireballs \cite{Rapp2013}. 
Currently, (3+1)-dimensional viscous hydrodynamics studies are standard tools for analyses and interpretation of heavy-ion collision results, and dileptons have also been shown to be sensitive to viscosities (shear and bulk) and to details of the initial 
conditions~\cite{Vujanovic2013,Vujanovic2016,Vujanovic2019}.

Recently, a great deal of attention has been devoted to the  
baryon-dense region of the QCD phase diagram~\cite{Savchuk2022}, 
which is being explored by the Beam Energy Scan (BES)
program at RHIC, 
and by the NA61/SHINE experiment at the CERN SPS.
The future NA60+ experiment is being proposed to measure 
the dimuon spectrum for baryon-rich systems~\cite{NA60plus,Scomparin2022}. Anticipating detailed comparisons with those and other experiments, we concentrate solely on the thermal contribution from deconfined quarks and gluons,
reporting on two aspects of this investigation: 
$i$) we generalize the NLO perturbative calculation of dilepton emission rates to non-zero baryon 
chemical potential $\mu_{\rm B}$
and
$ii$) we embed those newly derived dilepton rates in  
modern (3+1)-dimensional relativistic hydrodynamical simulations of heavy ion collisions. We focus our analysis on the invariant mass region
1\,GeV${\,<\,}M{\,<\,}3$\,GeV where QGP manifestations are expected to be prevalent \cite{Rapp:2014hha,Gale:2014dfa}. It will be seen that the presence of a nonzero $\mu_{\rm B}$ modifies the quark and antiquark
distributions in the QGP, and shifts the thermal masses that control the associated screening effects.
While the latter has been examined for real photons \cite{Gervais2012}, in the first part of this paper 
we will present new results away from the light cone.
This involves properly understanding how $\mu_{\rm B}$ enters the strict NLO 
computation, how it affects the so-called Landau-Pomeranchuk-Migdal (LPM) effect, 
and how to smoothly interpolate between the two regimes as originally
advocated in Ref.~\cite{Ghisoiu2014}. 
In addition, we also provide the dilepton yield's dependence on 
polarisation. 
The second part of our paper is devoted to integrating the new rates just discussed, using a realistic fluid dynamical simulation of the evolving medium. 

Our work is organized as follows: the next section lays out general considerations pertaining to the evaluation of lepton pair production rates, and also sets up the general formalism needed for fluid dynamical simulations in environments with a net baryon number. Section \ref{sec:pert} considers the different relativistic field theoretical aspects  required to compute the photon self-energy at NLO in QCD in the presence of baryons, and we present the new dilepton rates. The following section, Section \ref{sec:pheno}, discusses the phenomenology made possible by the measurement of lepton pairs at RHIC. We then summarize and conclude.

\section{Setup}

\subsection{Differential emission rate \label{sec:setup}}

To fix the notation, we denote the temperature by $T$, the quark chemical potential by $\mu$ and
the energy and momentum of the lepton pair by $\omega$ and $\bm k$ respectively.
In chemical equilibrium,
$\mu = \frac13 \mu_{\rm B}$ where $\mu_{\rm B}$ is 
the chemical potential associated with baryon number density $n_{\rm B}$.
For now, we consider $\omega$ and $k$ to be defined by the
local rest frame (LRF) of the plasma.
We use a metric signature such that
$K^2 \equiv K_\mu K^\mu = \omega^2 - {\bm k}^2$
is the invariant mass associated with the
dilepton's four-momentum $K_\mu=(\omega,\bm k)$.

The dilepton emission rate per unit volume, $\Gamma_{\ell \bar \ell}$, 
of an equilibrated QGP can be derived from the retarded photon self-energy, 
$\Pi_{\mu\nu}(\omega, \bm k)\,$.
It is the imaginary part of $\Pi_{\mu\nu}$ that defines a spectral function, 
which is thus {\em fully} characterised by 
two polarisations 
$$
\rho_{\mu\nu} \equiv  {\rm Im}\big[ \Pi_{\mu\nu} \big]
 = 
\mathds{P}^{\rm L}_{\mu\nu} \, \rho_{\rm L}
 + 
\mathds{P}^{\rm T}_{\mu\nu} \, \rho_{\rm T}  ,
$$
where we  may take, \eg,  the following projectors in the fluid's LRF~\cite{Weldon1990}, 
\bea
\mathds{P}^{\rm L}_{\mu\nu} 
 = 
  \frac{1}{K^2} 
\l( \begin{array}{cc}
  \bm k^2 & \omega \bm k \\
  \omega \bm k     &  \omega^2 \hat{k}_i \hat{k}_j
\end{array} \r),
\
\mathds{P}^{\rm T}_{\mu\nu} 
=  
  \l( \begin{array}{cc}
  0 & 0 \\  0  &  (\delta_{ij} - \hat{k}_i \hat{k}_j)
\end{array} \r) \, . \!\!\!\!\!\!\!\! \nonu\\
\label{projectors} 
\eea

Of course, real photons satisfy $\omega = k \equiv |\bm k|\,$, 
whereas dileptons are issued forth by virtual photons with 
enough invariant mass $M \equiv \sqrt{ \omega^2 - \bm k^2}$ to create 
two particles, each of mass $m_\ell\,$, \ie\ $M > 2m_{\ell}\,$.
If the final state leptons carry four-momenta $P_+$ and $P_-$ (so that $K=P_++P_-$), 
 we define the associated leptonic tensor by 
$L^{\mu\nu} = P_+^\mu P_-^\nu + P_+^\nu P_-^\mu - g^{\mu\nu}( P_+ \cdot P_- + m_\ell^2)\,$.
Then, to leading order in the electromagnetic fine-structure constant 
$\alpha_{\rm em}$ and neglecting quark masses, the emission rate equals~\cite{Gale1990}
\begin{widetext}
\bea 
 E_+ E_- \frac{\dd \Gamma_{\ell \bar \ell}}{\dd^3 {\bm p}_+ \dd^3 {\bm p}_-} 
= 
\frac{ \nB (\omega) }{2\pi^4 \, M^4}
\ \bigg\{  \alpha_{\rm em}^2 \, \sum_{i=1}^{\Nf} Q_i^2 \bigg\}
\,
B \Big( \frac{m_\ell^2}{M^2} \Big) 
\  L^{\mu\nu}\rho_{\mu\nu} (\omega, k).
\label{1}
\eea 
\end{widetext}
Here the Bose distribution function is
$\nB(\omega)\,$, the quark charge-fractions are $Q_i$ (in units of the electron charge),
and the kinematic factor to produce the pair of leptons is 
$B(x) \equiv (1 + 2x) \Theta( 1 -4x ) \sqrt{ 1 -4x }\,$. 

Equation \eq{1} makes it clear that the photon spectral function is the central
object of interest.\footnote{%
 Equation \eq{1} is true to all orders in $\alpha_s=g^2/(4\pi)$~\cite{Bodeker:2015exa}.
}
Being the imaginary part of a retarded self energy, with averages taken in a thermal ensemble, $\rho_{\mu\nu}$
is intimately related to the Euclidean correlator---an
object that can be estimated\footnote{Information about $\rho_{\mu\nu}$ for {\em all} $\omega$ is needed to calculate the Euclidean correlator, while the dilepton rate only depends on the spectral function in the timelike region.} 
from continuum-extrapolated lattice data~\cite{Blaizot:2005mj}.
The resummed NLO spectral function holds up to scrutiny when tested against lattice results for 
$\Nf = \{ 0,2 \}$~\cite{Jackson2019,Jackson2022} and $\Nf=3$~\cite{Bala2022}
(all at zero $\mu_{\rm B}$),
which motivates using the associated perturbative QCD rates to compute the actual yield from simulations.

The theoretical evaluation of the dilepton rate has a long history, with 
time and persistence having clarified most of the important physics issues.
Initially, calculations assumed that the photon was at rest,\ \ie $k=0$, and $\omega \sim T$ 
which allowed the current-current correlator to be determined from the 
two-loop photon self energy~\cite{Baier1988,Gabellini1989,Altherr1989}. 
It was soon recognised that such an approach breaks down for $\omega \lsim \sqrt{\alpha_s}\, T$, 
where hard thermal loops (HTLs)~\cite{Braaten:1989mz} need to be 
resummed to account for screening and Landau damping~\cite{Braaten1990,Moore2006}.
The same techniques were used to derive the photon rate ($\omega=k$), which is proportional to $2\rho_{\rm T}$
and has the parametric behaviour 
$\omega\, d\Gamma_\gamma/d^3\bm k \sim \alpha_{\rm em} \, \alpha_s \log (1/\alpha_s) T^2$~\cite{Kapusta1991,Baier1991}. 
Spacelike virtualities ($\omega < k$) were only recently discussed in the literature, 
when first needed to connect with the Euclidean correlator---physically this 
concerns the somewhat academic case of deep inelastic scattering on a QGP target~\cite{harvey_DIS}. 
Whether the photon point is approached from above or below $K^2\sim \alpha_s T^2$ (but keeping $\omega \sim T$) 
naively higher-order diagrams must be incorporated~\cite{Aurenche2002,Aurenche:2002pc}. 
In addition to HTL resummation, certain multiple scatterings are also mandated by 
collinear enhancement (the LPM effect) to rigorously give the 
leading\footnote{%
  In our convention, the LO spectral function refers to the QED result for $\rho_{\mu\nu}$ and the NLO
  spectral function includes the terms  proportional to $\alpha_s\,$.
  Papers that focus on the photon rate will refer to the latter as the `LO' result because 
  for $\omega \simeq k$ the process $q \bar q \to \gamma^\star$ is kinematically suppressed and the
  spectral function starts with the QCD `corrections.'
  Relative ${\cal O}(\sqrt{\alpha_s})$ terms in this limit have also been established~\cite{Ghiglieri2013,Ghiglieri2014}. 
  } QCD corrections~\cite{Arnold2001ba,Arnold2001ms,Aurenche2002}. 
The opposite limit, namely $M \gg T$, can be studied independently via the operator product expansion (OPE)~\cite{ope}. 

A few extra considerations are needed to complete the calculation of $\rho_{\mu\nu}$ for $\muB \neq 0\,$, 
which we will describe in Sec.~\ref{sec:pert}, but no conceptual obstacles remain.

\subsection{QGP spacetime evolution}\label{sec:hydro}

Heavy-ion collisions are commonly described by multistage frameworks, which consist of a sequence of physical models to describe the evolution stages of the created systems \cite{Shen:2020jwv,iebe}. The hydrodynamic description of the deconfined QGP phase is the core of such frameworks. It describes the spacetime evolution of a system which
is near local (chemical and thermal) equilibrium, with conservation of energy and momentum and that of baryon charge density \cite{Denicol2018,Du:2019obx}:
\be
    \pd_\mu T^{\mu\nu}
    \ =\ 0
    \,,\quad 
    \pd_\mu J_{\rm B}^\mu \ =\ 0\,,
    \label{conservation}
\ee
where $T^{\mu\nu}$ is the energy-momentum tensor and $J_{\rm B}^\mu$ the net baryon charge current.
In practice, the system of Eqs.~\eq{conservation} is solved numerically
in coordinates where $\tau = \sqrt{t^2-z^2}$ is the proper time and
$\eta_s = \frac12 \log{\frac{t+z}{t-z}}$ is the spacetime rapidity (with $z$ being the Cartesian coordinate parallel to the beam direction).

We define the energy density $e$ in the Landau frame, i.e. 
$u_\mu T^{\mu\nu} \equiv e \, u^\nu$, so that 
\bea
T^{\mu\nu} &=&
e \, u^\mu u^\nu
-
\big(\,p + \Pi\,\big)\,
\Delta^{\mu\nu} 
+
\pi^{\mu\nu}\, ,
\eea
where $\Delta^{\mu\nu} \equiv g^{\mu\nu} - u^\mu u^\nu$, $u^\mu$ is the fluid four-velocity, $p+\Pi$ is the usual-plus-bulk viscous pressure, and $\pi^{\mu\nu}$
is the shear stress tensor.
The baryon current reads
\bea
J_{\rm B}^\mu &=&
n_{\rm B} \, u^\mu + q^\mu \, ,
\eea
 with $n_{\rm B}$ the local baryon number density and $q^\mu$ the associated diffusion current~\cite{Denicol2018}. To close the system of continuity equations for $T^{\mu\nu}$ and $J_{\rm B}^\mu\,$,
one also requires the equation of state (EoS), \ie\ $p=p(e,n_{\rm B})\,$, and
a set of equations of motion for each of 
the dissipative currents $\Pi$, $\pi^{\mu\nu}$ and $q^\mu$ \cite{Denicol2012}.

The pre-hydrodynamic stage becomes more complicated and important at lower collisional energies,
where the Lorentz contraction of the nucleus is not sufficient to disregard
the time it takes the two nuclei to entirely pass through one another.
This is the case for the NA60+ experiments, which operate in the SPS energy
range 5\,GeV${\,<\,}\sqrt{\sNN}{\,<\,}$17\,GeV, and the Beam Energy Scan program at RHIC.
We follow the initialisation procedure 
of Refs.~\cite{Denicol2018, Du:2023gnv}, which used an averaged Monte Carlo Glauber sampling for the initial
conditions as input for hydrodynamic simulations starting at 
$\tau_0$. 
The hydrodynamical evolution, with dissipative corrections from $\pi^{\mu\nu}$ and $q^\mu$~\cite{Du:2023gnv}, is simulated using the MUSIC fluid-dynamical approach~\cite{music1,music2}. We use the 
equation of state {\sc neos-b}~\cite{Monnai2019} which neglects strangeness and electric charge chemical potentials.

\begin{figure}[!tbp]
    \centering
    \includegraphics[width= 0.9\linewidth]{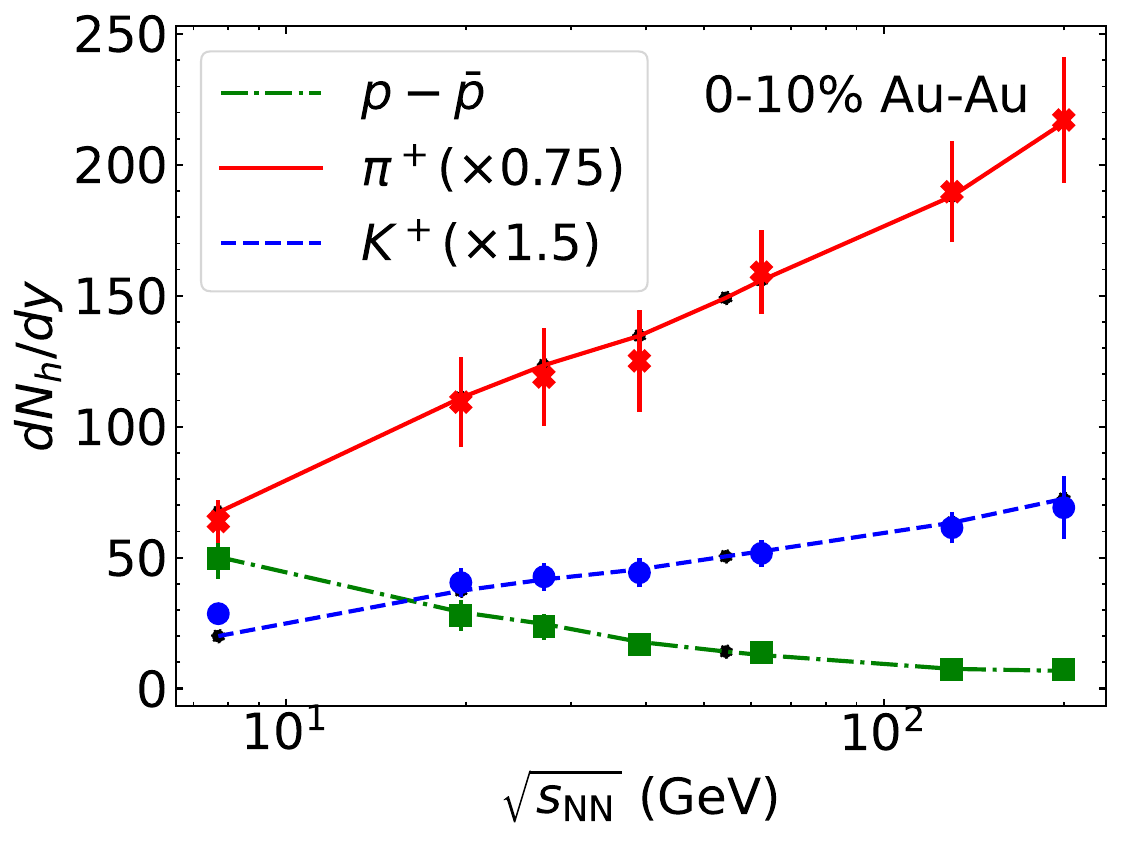}
    \caption{Identified hadron yields at midrapidity for 0-10\% Au+Au collisions at eight beam energies from 7.7 to 200 GeV. The markers show the STAR measurements for net protons $p$-$\bar p$ (green square), pions $\pi^+$ (red cross), and kaons $K^+$ (blue circle) and the lines connect modeling results shown by black dots. }
    \label{fig:BES_hadron_yields}
\end{figure}

\begin{figure}[!tbp]
    \centering
    \includegraphics[width=0.9\linewidth]{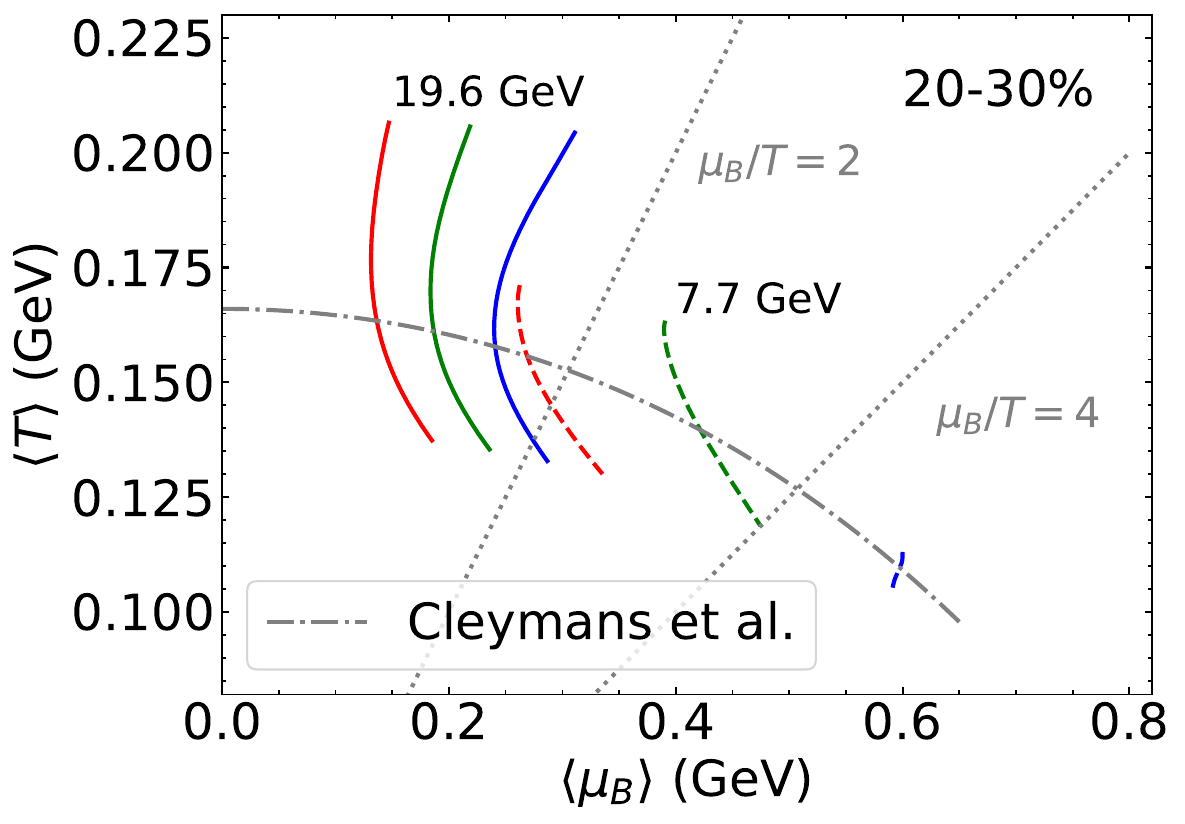}
    \caption{Hydrodynamic expansion trajectories in the phase diagram within three distinct spacetime rapidity windows $(-0.5,\,0.5)\,,$ $(0.5,\,1.0)\,,$ and $(1.0,\,1.5)\,$ (from left to right curves) for 20-30\% Au+Au collisions at 7.7 GeV (dashed lines) and 19.6 GeV (solid lines). The two dotted lines correspond to $\mu_{\rm B}/T{\,=\,}$2 and 4, respectively. The dot-dashed line represents the chemical freeze-out line from Ref. \cite{Cleymans:2005xv}.}
    \label{fig:phase_diagram_traj}
\end{figure}

\bigskip

The hydrodynamics evolves until $e_{\rm fo} = 0.26$~GeV/fm$^3$, 
to allow for a transient hot hadronic gas which is slightly cooler than the QGP.
Once the hydrodynamic stage is complete, a freeze-out surface is generated and read into iS3D, 
a numerical simulation that uses the Cooper-Frye formalism to convert $T^{\mu\nu}$ 
into particles while conserving energy and momentum, and baryon charge. 
The subsequent hadronic re-scattering and resonance decays of the dilute phase 
are determined from UrQMD~\cite{Bass1998,Bleicher1999}. The initial conditions are tuned so that the final hadron yields match the experimental measurements at midrapidity (see Fig.~\ref{fig:BES_hadron_yields}) and away from midrapidity if available (see Ref.~\cite{Du:2023gnv}). 
Figure~\ref{fig:phase_diagram_traj} depicts the hydrodynamic expansion trajectories ($\muB, T$) in the phase diagram across three distinct spacetime rapidity windows for the two lowest beam energies. 
As shown in the figure, the created systems achieve higher chemical potentials and lower initial temperatures as the beam energy decreases, leading to shorter expansion trajectories, as observed at $7.7$~GeV away from midrapidity.
Notably, it reveals that, even at a beam energy of $7.7$~GeV, a significant portion of fluid cells satisfies $\muB/T\lsim~3$ within $|\eta_s|<1$.

\section{Perturbative results, $\bm \mu \, \bm \neq\, \bm 0$\label{sec:pert}}

Before getting into the details, let us reiterate that our study is dedicated to the {\em equilibrium} thermal dilepton rate, \ie\ we do not take into account any departure of the quark or gluon distribution functions from thermal and chemical equilibrium, as would occur in the presence of viscosity and/or diffusion.\footnote{%
As mentioned in Sec.~\ref{sec:hydro}, corresponding effects are included in the hydrodynamic evolution.
}
Although it has recently become clear how such corrections may be implemented, even when considering LPM-type diagrams~\cite{Hauksson:2017udm,Vujanovic2019}, 
 we concentrate here on the changes brought about by including NLO and finite $\muB$ effects and leave off-equilibrium considerations for future work.

According to the convention in Eq.~\eq{projectors}, 
the longitudinal (L) and transverse (T) components can be expressed by
\bea
\rho_{\rm L}  =  - \frac{K^2}{k^2} \rho_{00} \ ,
& & 
\rho_{\rm T} \ = \frac12 \Big( \rho_{\mu}^{\ \mu} + \frac{K^2}{k^2} \rho_{00} \Big)
\, .
\eea
We will denote the full trace by $\rho_{\rm V} \equiv \rho_{\mu}^{\ \mu} = 2\, \rho_{\rm T} + \rho_{\rm L}\,$, which is needed for the dilepton rate.
For the purpose of displaying formulas,
it is simpler to focus on $\rho_{\rm V}$ and $\rho_{00}\,$.

For the spectral functions, the LO results are
\vspace{-2mm} 
\bea
 \rho_{\rm V} & = &
  \frac{ \Nc K^2 }{4 \pi k }
  \Big\{
    T \sum_{\nu \,=\, \pm \mu}
    \Big[\, \lnf\big(\kp-\nu\big) - \lnf\big(|\km|-\nu\big)\,\Big]\nonumber \\
 &+& k \, \theta(\km^{ })
  \Big\}
\;, \label{rhoV_LO}\\
  \rho_{00} & = &
  \frac{ -\Nc }{12 \pi k }
  \Big\{ 
    12 T^3 \!\!
    \sum_{\nu \,=\, \pm \mu}
    \Big[\, \ltf\big(\kp-\nu\big) - \ltf\big(|\km|-\nu\big) \,\Big]
    \label{rho00_LO}
\nonumber \\
 &+&  \!
  6 k T^2 \!\!\!
    \sum_{\nu \,=\, \pm \mu}
 \Big[ \lif\big(\kp-\nu\big) + \sign(\km) \lif\big(|\km|-\nu\big)\Big]\nonumber \\
 &+&  k^3 \, \theta(\km) 
  \Big\} 
 \;, 
\eea
\vspace{-1mm} 

\noindent where 
{$\theta$ is the Heaviside step function and\ 
we have defined 
the polylogarithms 
\bea
 \lnf(x)  &\equiv&  \log \Big( 1 + e^{-x/T} \Big), \ 
 \lif(x) \;\, \equiv \; \mbox{Li}^{ }_2 \Big(-e^{-x/T}\Big),\nonumber \\  
 & &\ltf(x) \;\, \equiv \; \mbox{Li}^{ }_3 \Big(-e^{-x/T}\Big)
 \;. \label{polylogs}
\eea
The function $\rho_{\rm V}$ (for $\mu\neq 0$) was previously determined in Ref.~\cite{Dumitru1993} for $\omega > k\,$.
However, equation \eq{rhoV_LO} and \eq{rho00_LO} are more general because
 they also hold for $\omega < k\,$.
Setting $\mu=0$ reproduces Eq.~(2.4) from Ref.~\cite{Jackson2019}.

At NLO, the underlying spectral function can be reduced to a set of elementary 
`master integrals' 
(many of which were studied for $\omega >k$ in \cite{Laine2013vpa}),
providing a way to divide and conquer the otherwise very tedious calculation.
The evaluation involves such integrals, whose imaginary
parts are uniformly defined by
\begin{widetext}
\bea
\rho_{abcde}^{(m,n)}(\omega,k)
&\equiv &
\Im \bigg[ \ 
\displaystyle
\sumint{\, P,Q} \, 
\frac{p_0^m \, q_0^n}{
  [P^2]^a \, [Q^2]^b \, [(K-P-Q)^2]^c \, [(K-P)^2]^d \, [(K-Q)^2]^e
}
\ \bigg]_{k_0 \to \omega + i0^+}
\, .
\label{I def}
\eea
This spectral function corresponds to
the generic two-loop topology for a self energy
of external momentum $K\,$.
In the sum-integrals,\footnote{%
To be crystal clear, with $d=D-1$, the sum-integrals are:
$$
\sumint{\, P} = \int_{\bm p} \, T \sum_{p_0} \ ; \qquad
\int_{\bm p} = \l( \frac{e^\gamma \bar \mu^2}{4\pi} \r)^\epsilon 
\int\!\! \frac{\dd^d p}{(2\pi)^d} \, .
$$
}
$P$ and $Q$ are fermionic momenta with
$p_0 = i (2x+1)\pi T + \mu$ and
$q_0 = i (2y+1)\pi T - \mu$ (where $x,y \in \mathbb{Z}$).

Turning now to the specific diagrams from Fig.~\ref{fig:nlo}, which we
evaluate in
$D=4-2\varepsilon$ spacetime dimensions, 
the result is a linear combination of the functions of the type~\eq{I def}.
More specifically, the ${\cal O}(g^2)$ contributions read
\bea
 \left.\rho_{\rm V}\right|^{(g^2)}_{\rm NLO} & = &  
\label{masters, mn} 
4(D-2)
g^2 \CF \Nc   \bigg\{ 
(D-2)\frac{K^2}2
\Big( 
  \rho_{11020}^{(0,0)} + \rho_{11002}^{(0,0)} 
- \rho_{10120}^{(0,0)} - \rho_{01102}^{(0,0)} \Big)
+ \rho_{11010}^{(0,0)} \nonu \\ 
&+& \rho_{11001}^{(0,0)} 
+ (4-D) \, \rho_{11100}^{(0,0)} 
 + 2 \frac{K^2}{k^2} \rho_{11011}^{(1,1)}
- \tfrac12 K^2 
\Big(\, \frac{\omega^2}{k^2} + 7-D \,\Big) 
\rho_{11011}^{(0,0)} \\
&-&  \tfrac12 (D-2) \Big( \rho_{1111(-1)}^{(0,0)} + \rho_{111(-1)1}^{(0,0)}  \Big)
+ 2K^2  \Big( \rho_{11110}^{(0,0)} + \rho_{11101}^{(0,0)} \Big)
- K^4 \rho_{11111}^{(0,0)} 
\ \bigg\}  \ ,\nonu \\
\left.\rho_{00}\right|^{(g^2)}_{\rm NLO} & = &
\label{masters, 00}
4 g^2 \CF \Nc  \bigg\{ 
 \tfrac12 (D-2)\Big(\rho_{10110}^{(0,0)} + \rho_{01101}^{(0,0)}  \Big)
+  (4-D) \rho_{11100}^{(0,0)} 
+ \big[ 
\tfrac12 D \, K^2 - \omega^2 + 3k^2
\big] \, \rho_{11011}^{(0,0)} \nonu \\
&-&  \tfrac12 (D-2) \Big( \, 4\rho_{11011}^{(1,1)} 
+ \rho_{1111(-1)}^{(0,0)} + \rho_{111(-1)1}^{(0,0)} \Big) 
+ 
 2\big[\,K^2 -(4-D)\, \omega^2  \,\big]
 \Big( \rho_{11110}^{(0,0)}  + \rho_{11101}^{(0,0)} \Big) \nonu \\ 
&+& 2(4-D) \, \omega 
 \Big( \rho_{11110}^{(1,0)} + \rho_{11101}^{(0,1)} \Big)
- 2(D-2) \omega \Big( \rho_{11110}^{(0,1)} + \rho_{11101}^{(1,0)} \Big) \nonu \\
&+&
  \big[(D-3) \omega^2 + k^2\big] \,K^2 \rho_{11111}^{(0,0)} 
+ 2(4-D) \, K^2 \rho_{11111}^{(1,1)} 
- (D-2) K^2\Big( \rho_{11111}^{(2,0)} + \rho_{11111}^{(0,2)} \Big)
\,\bigg\} \ . 
\eea
\end{widetext}

Above, the limit $D \to 4$ is implied because
several of the master integrals have $1/\varepsilon$-divergences stemming from
their vacuum parts.
These all cancel when combined in the expressions
above, but some care is required to verify that the result
is finite~\cite{Jackson2019a}.
In order to arrive at Eqs.~\eq{masters, mn} and \eq{masters, 00},
certain identities have been used to minimize the number of master integrals required.
The same `minimal basis' of functions was studied (for $\mu=0$) in
Refs.~\cite{Laine2013vpa,Laine:2013vma,Jackson2019a}.
It is relatively straight forward to generalise those results
to finite baryon densities; see Appendix~\ref{app:A}.
Note that both Eq.~\eq{masters, mn} and Eq.~\eq{masters, 00}
involve linear combinations which are symmetric in the simultaneous
exchanges:
$a \leftrightarrow b\,$, $d \leftrightarrow e$ and $m \leftrightarrow n\,$ 
for the master integrals.
Consequently, the result will be unchanged by $\mu \to - \mu\,$.
In the case where $\mu = 0\,$, using 
$\rho_{abcde}^{(m,n)}
= \rho_{baced}^{(n,m)}$ leads to the same decomposition as in Ref.~\cite{Jackson2019a},
as expected.

\begin{figure}[!tbp]
  \includegraphics[width=\linewidth]{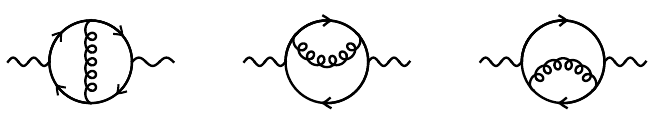}
  \caption{\label{fig:nlo} 
  Strict two-loop contributions to the photon self-energy in QCD. 
  (The 2nd and 3rd diagrams are distinct for $\mu\neq 0\,$.)
  }
\end{figure}

\subsection{Resummation in the LPM regime}
The strict perturbative expansion is valid as long as $K^2 \gsim T^2$.
In the opposite limit, \ie\ $K^2 \to 0^\pm$, 
resummation is needed to obtain physically meaningful results.
Thermal screening is the essential requirement, 
but the LPM effect also needs to be included for completeness.
This has been well established at $\mu=0$~\cite{Arnold2001ba,Arnold2001ms,Aurenche2002}.
At non-zero baryochemical potential, results exist for the photon rate, $\omega = k\,$,
which was considered in Ref.~\cite{Traxler1994} and then the LPM effect was incorporated by 
Ref.~\cite{Gervais2012}.
Here we extend those results to $\omega \lg k\,$ (as well as studying $\rho_{\rm T}$ and $\rho_{\rm L}$ 
separately).

Two important scales in the problem are modified by the presence of a chemical potential: 
the Debye mass $m_D$ and the `asymptotic' quark mass $m_\infty$,
{\em viz.},
\bea
m_D^2   &\equiv& 
g^2 \bigg[
\Big(\tfrac12 \Nf + \Nc \Big)
\frac{T^2}3 
+
\Nf \, \frac{\mu^2}{2\pi^2} \bigg] \, ,\nonu \\
m_\infty^2 &\equiv &
g^2 \, \frac{\CF}4 \bigg( T^2 + \frac{\mu^2}{\pi^2} \bigg) \; .
\label{thermal masses}
\eea

For $K^2 \to 0^\pm\,$, the spectral function needs to be 
resummed in order to account for the LPM effect,
with $m_D$ and $m_\infty$ being the key ingredients.
In practice, kinematic approximations are
invoked to perform the resummation to all orders
for diagrams of the type shown in Fig.~\ref{fig:lpm}.
Specifically, most formulations set up a two dimensional
effective kinetic description in the transverse plane.
Following the usual approach, we can express the spectral functions by
\bea
 \left. \rho^{ }_{00} \right|^{\rm full}_{\rm LPM} 
 & \equiv &  
 - \frac{\Nc}{\pi}
 \int_{-\infty}^{\infty} \! \dd\epsilon \, 
 \bigl[ 1-\nF(\epsilon-\mu)-\nF(\omega-\epsilon+\mu) \bigr] \nonu\\
 &\times& \int_{\bm p_\perp}
   \! \im [g({\bm p}_\perp)] \ ,
 \label{rho 00 LPM} \hspace*{5mm}\\
 \left. \rho_{\rm T} \right|^{\rm full}_{\rm LPM} 
 & \equiv &  
 - \frac{\Nc}{\pi}
 \int_{-\infty}^{\infty} \! \dd\epsilon \, 
 \frac{
 1-\nF(\epsilon-\mu)-\nF(\omega-\epsilon+\mu) 
 }{4 \epsilon^2 ( \omega-\epsilon)^2} \nonu \\
&\times& \big[\omega^2 - 2 \epsilon(\omega-\epsilon)\big]
 \int_{\bm p_\perp}
   \! \re [{\bm p}_{\perp}\cdot {\bm f}({\bm p}_\perp)] 
 \, , 
 \label{rho T LPM} 
\eea
where the quantities $g$ and $\bm f$ are functions of the transverse momentum coordinate $\bm p_\perp$
and satisfy the following integral equations\footnote{%
  A factor of $-i \sqrt{|\epsilon(\omega-\epsilon)|}$ is absorbed into the definition of $g$, 
  compared with \cite{Aurenche:2002pc}. 
  Our normalisation convention for $g$ and $\bm f$ matches Ref.~\cite{Ghisoiu2014}.
  }
\bea
2 \, i &=& i \, \delta E \, g(\bm p_\perp)  \label{integral eq g}\\
&+& g^2 \CF T \int_{\bm q_\perp} {\cal C}(\bm q_\perp) 
\big[ g(\bm p_\perp) - g(\bm p_\perp - \bm q_\perp) \big] \, ,  \nonu \\
2 \, \bm p_\perp &=& i \, \delta E \, {\bm f}(\bm p_\perp) \label{integral eq f}\\
&+& g^2 \CF T \int_{\bm q_\perp} {\cal C}(\bm q_\perp) 
\big[ {\bm f}(\bm p_\perp) - {\bm f}(\bm p_\perp - \bm q_\perp) \big] \, . \nonu
\eea
The required  collisional operator ${\cal C}(\bm q_\perp) = \frac1{q_\perp^2} - \frac1{q_\perp^2 + m_D^2}$ was found
in Ref.~\cite{Aurenche:2002pd}.
On the right hand side of Eqs.~\eq{integral eq g} and \eq{integral eq f}, $\delta E$ denotes an energy difference
which accounts for the relative phase of the emitted radiation in the amplitude and conjugate amplitude.
We shall explicitly define $\delta E$ shortly.

\begin{figure}[t]
  \includegraphics[scale=.85]{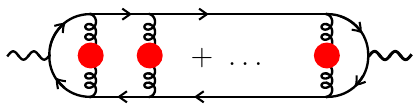}
  \caption{\label{fig:lpm} 
  Class of diagrams that contribute in the LPM regime. 
  Gluons have HTL resummed propagators and quarks are `hard,' 
  i.e. propagate with their asymptotic thermal mass.
  }
\end{figure}

It is convenient to solve the above integral equations in coordinate space,
which leads to the two dimensional Schr\"{o}dinger equations:
\bea
 \bigl( \hat{H} + i 0^+_{ }\bigr) g({\bm y}) &=& \delta^{(2)}({\bm y})
 \;, \nonu \\
 \bigl( \hat{H} + i 0^+_{ }\bigr) \vec{f}(\vec{y}) &=& -\nabla^{ }_{\perp} \delta^{(2)}(\vec{y}) 
 \;,
\ee
where $\bm y$ is a transverse coordinate.
The boundary conditions are such that $g$ and $\bm f$ should be 
regular at the origin, where they are needed for 
Eqs.~\eq{rho 00 LPM} and \eq{rho T LPM}. 
In practice, we construct them from the $S$ and $P$-wave solutions to the 
corresponding homogeneous differential equation as described in Ref.~\cite{Ghisoiu2014}.
The operator $\hat{H} = \delta E + i V$ acts in the plane transverse to light-like propagation,
with the potential
\be
V 
&=&
   g^2 \CF T
   \int_{\bm q_\perp}\!\!\!
  \bigl( 1 - e^{i \vec{y} \cdot \vec{q}_\perp }\bigr)
  {\cal C}( {\bm q}_\perp ) \, .
 \label{hatH}
\ee
Before giving $\delta E\,$, let us recall that the LPM resummation assumes $\omega \gg g T$
and $K^2 = \omega^2 - k^2 \ll \omega^2\,$.
Hence, certain terms may be exchanged, \eg $k-\omega$ is equivalent to
 $-K^2/(2\omega)$ at this level of accuracy.
The ambiguity starts to matter (numerically) when going beyond the presupposed strict kinematics.
One can, however, mitigate this effect for practical purposes by maintaining a certain cancellation
between the transverse and longitudinal polarisations that occurs at the Born level~\cite{Aurenche2002,Ghisoiu2014}.
We adopt this standard procedure, by 
multiplying $\rho_{00}$ by $K^2/\omega^2$ in \eq{rho 00 LPM} to obtain the longitudinal polarisation
and to express the energy shift by
\bea
\delta E &=&
- \, \frac{K^2}{2\omega}
+ \frac12 \Big( \, \frac1{\epsilon} + \frac1{\omega-\epsilon} \, \Big)
\big( m_\infty^2 - \nabla_\perp^2 \big)  \, .
\label{hamiltonian}
\eea
The full spectral function can then be determined and 
the two projections $\rho^{ }_i$, with $i={\rm T},{\rm L}$, are given by 
\bea
 \left. \rho^{ }_i \right|^{\rm full}_{\rm LPM} 
 & \equiv &  
 - \frac{\Nc}{\pi}
 \int_{-\infty}^{\infty} \! \dd\epsilon \, 
 \bigl[ 1-\nF^{ }(\epsilon-\mu)-\nF^{ }(\omega-\epsilon+\mu) \bigr]\nonu \\
 &\times & \!\!\!
 \lim_{{\bm y}\to {\bm 0}} \, 
 \biggl\{ \ 
   \delta^{ }_{i,{\rm L}}
   \frac{K^2}{\omega^2}
   \im [g({\bm y})] \nonu \\
  & & \!\!\!\hspace{0.6cm}\, + \, 
   \frac{ \delta^{ }_{i,{\rm T}} }{4}
   \Big( \,
   \frac{1}{\epsilon^2} + 
   \frac{1}{(\omega-\epsilon)^2}
   \, \Big) 
   \im [\nabla^{ }_{\perp}\cdot {\bm f}({\bm y})]  
 \, \biggr\} 
 \, . \nonu\\
 \label{final_from_LPM} \hspace*{5mm}
\eea

\subsection{How to combine NLO and LPM}

As mentioned, the LPM regime relies on certain
kinematic simplifications that only hold near the light cone.
The strict NLO results involve general kinematics, but
are plagued by an unphysical divergence as $K^2 \to 0^\pm$
which is precisely why resummation is needed there.
We now discuss how these two formulations can be combined
for `intermediate' $K^2$.

The 
resummed spectral functions 
($i \in \{{\rm V},{\rm T},{\rm L},\ldots\}$)
are defined as~\cite{Ghisoiu2014}\footnote{%
A more sophisticated way of interpolating
was suggested in Ref.~\cite{Ghiglieri2021},
and involves carefully modifying \eq{hamiltonian}.
}
\bea
 \rho_i |_{\rm NLO}^{\rm resummed}
 &\equiv& 
 \rho_i |_{\rm 1-loop}^{\rm strict}
 + 
 \rho_i |_{\rm 2-loop}^{\rm strict} \nonu\\
 &+& 
 \big( 
 \rho_i |_{\rm LPM}^{\rm full}
 - 
 \rho_i |_{\rm LPM}^{\rm expanded}
 \, \big) 
 \;, \label{resummation} 
\eea
where $\rho_i |_{\rm 1-loop}^{\rm strict}$
is determined from Eqs.~\eq{rhoV_LO} and \eq{rho00_LO},
$\rho_i |_{\rm 2-loop}^{\rm strict}$ is
determined from Eqs.~\eq{masters, mn} and \eq{masters, 00},
and $\rho_i |_{\rm LPM}^{\rm full}$ is 
given by Eq.~\eq{final_from_LPM}.
The 
quantity $\rho_i |_{\rm LPM}^{\rm expanded}$
is necessary for the `matching' procedure and will be discussed in the next paragraph.   
Equation~\eq{resummation} is illustrated for $k=2\pi T$ in Fig.~\ref{fig:spf},
demonstrating that it 
is both finite and continuous across the light cone,

\begin{figure}[t]
  \includegraphics[scale=.75]{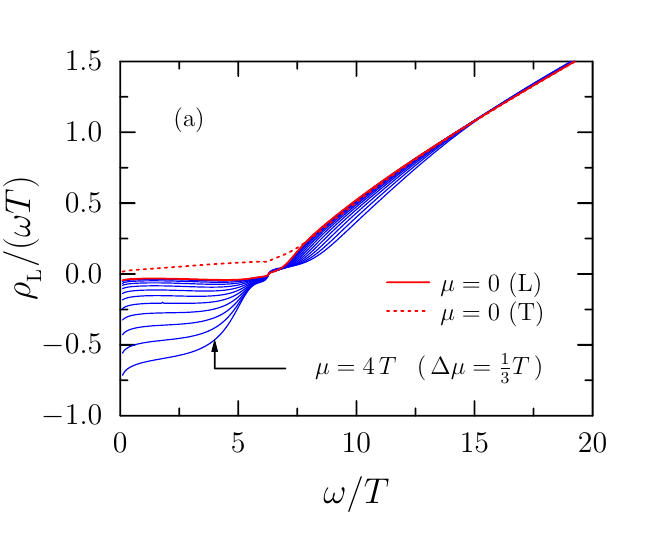}
  \vskip -6mm
  \includegraphics[scale=0.75]{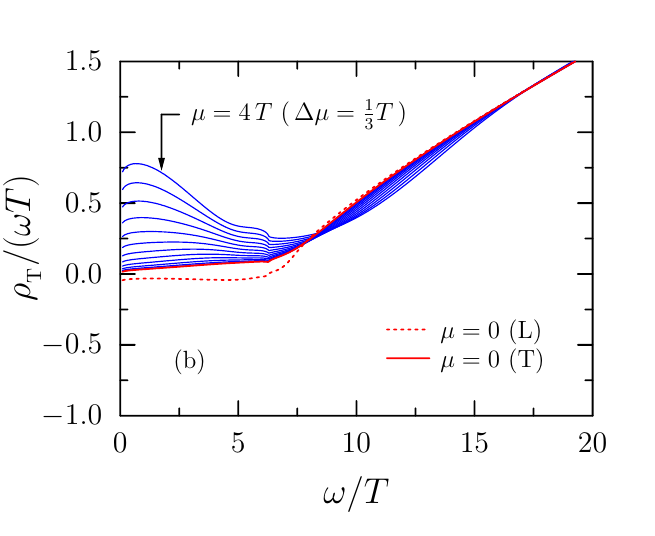}
  \vskip -3mm
  \caption{\label{fig:spf} 
  Transverse and longitudinal spectral functions, with $k= 2\pi\,T$ for a fixed
  value of the strong coupling $\alpha_s = 0.3\,$. The red curves in each
  figure depict $\rho_{\rm T,L}$ for zero chemical potential.
  The solid lines depict $\rho_{\rm L}$ in panel (a) 
  and $\rho_{\rm T}$ in panel (b). 
  The family of blue curves indicates the effect of varying 
  $\mu = 0, ..., 4\, T$
  in steps of $\Delta \mu = \frac13 T\,$.
  }
\end{figure}

In order to combine the LPM and NLO results, 
we need to `re-expand' the LPM results up to ${\cal O}(g^2)$
and remove double counting.
This is because the full LPM spectral function contains 
one and two-loop diagrams under simplifying assumptions,
which should rather be supplanted by the more correct LO and NLO results.
At zeroth order in $g$,  
the expressions become
\bea
 \rho_{\rm T} \big|_{\rm LPM}^{(g^0)}
 &=&  
 2 \, \frac{\Nc K^2}{\omega^2} \, 
 \big\langle \epsilon^2 + (\omega-\epsilon)^2 \big\rangle
 \;,\nonumber \\
 \rho_{\rm L}^{ } \big|_{\rm LPM}^{(g^0)} 
 & = &
 8\, \frac{\Nc K^2}{\omega^2} \,  
 \big\langle \epsilon (\omega - \epsilon ) \big\rangle
 \;, \label{LPM_LO}
\eea
where the angle brackets mean 
\bea
\big\langle ... \big\rangle
&\equiv&
 \frac1{16\pi \omega} \biggl\{ 
   \theta(K^2) \int_0^{ \omega^{ }} \!\! \dd\epsilon 
 - \theta(-K^2)
 \biggl[  \int_{-\infty}^0 \!\! + \int_{\omega^{ }}^{\infty} \biggr] 
 \, \dd\epsilon
 \biggr\} \nonu\\
 &\times &
 \bigl[ \nF^{ }(\epsilon - \omega -\mu) - \nF^{ }(\epsilon-\mu) \bigr]
 \, (...) 
 \, . 
\eea
These results should be compatible with Eqs.~\eq{rhoV_LO} and \eq{rho00_LO}
in the small-$K^2$ limit, as confirmed by
carrying out the explicit integration.
The expressions can be represented in terms of
\bea
 \langle \omega^2 \rangle
 &=&
 \frac1{16\pi} \biggl\{ 
 \theta(K^2) \, 
 \omega^2 \nonu\\
 &+& \omega T
 \sum_{\nu=\pm\mu} \bigl[ \lnf(\omega-\nu) - \lnf(\nu) \bigr]
 \biggr\}
 \, , \label{I_1}
 \eea
 \bea
 \langle \epsilon(\omega-\epsilon) \rangle
 &=& 
 \frac1{32\pi} \biggl\{ 
 \theta(K^2) \, 
 \frac{\omega^2}{3} \nonu\\
 &+& 2 T^2  \!\!
 \sum_{\nu=\pm\mu} \bigl[ \lif(\omega-\nu) + \sign(K^2)\, \lif(\nu) \bigr]
 \nonu\\
 &+& 4 \frac{T^3}{\omega}  \!\!
 \sum_{\nu=\pm\mu} \bigl[ \ltf(\omega-\nu) - \ltf(\nu) \bigr]
 \biggr\}
 \, . 
 \label{I_2} 
\eea
The next corrections are of order $g^2$,
and are proportional to the asymptotic mass $m_\infty^2\,$.
As in the $\mu=0$ case~\cite{Jackson2019}, 
for $\rho_{\rm L}$ there is no such contribution for non-zero $\mu$:
\bea 
 \rho_{\rm L} \big|^{(g^2)}_{\rm LPM} &=& 0 \, .
\eea
However, the transverse spectral function $\rho_{\rm T}$ contains a
logarithmic divergence, plus a finite part, namely
\bea
 \rho_{\rm T} \big|^{(g^2)}_{\rm LPM} 
 & = & 
 2 \Nc^{ } m_\infty^2 
 \Biggl\{ 
  \bigg\langle
 2 - 
  \frac{\omega}{\epsilon_-}
  -
  \frac{\omega}{(\omega-\epsilon)_+}
 \bigg\rangle
 \nonu\\
 &+&
 \frac{1}{16\pi}
   \biggl[ 1 - \nF(\omega-\mu) - \nF(\omega+\mu) \biggr] \nonu \\
   &\times& \,
   \biggl( \log\biggl| \frac{m_\infty^2}{K^2}  \biggr| - 1 \biggr)
 \Biggr\} 
 \, . 
   \label{LPM T}
\eea
We have introduced a convenient ``$\pm$ notation'' to regularize the
two singularities appearing above,
\bea
\int \dd\epsilon \, \frac{f(\epsilon)}{\epsilon_-}
&\equiv& 
\int \dd\epsilon \, \frac{f(\epsilon)-f(0)}{\epsilon} \, ,\\
\int \dd\epsilon \, \frac{f(\epsilon)}{(\omega-\epsilon)_+}
&\equiv& 
\int \dd\epsilon \, \frac{f(\epsilon)-f(\omega)}{\omega-\epsilon} \, .
\eea

For the resummation in Eq.~\eq{resummation} to be valid, it 
is crucial that the $\log |K^2|$ term, evident in \eq{LPM T},
is precisely cancelled by the NLO computation \eq{masters, mn}.
This is shown analytically in Appendix~\ref{app:B}
and is self-evident in the numerical results plotted in Fig.~\ref{fig:spf}, which shows the longitudinal and transverse spectral functions as a function of the scaled energy, $\omega/T$, for a fixed momentum $k/T = 2 \pi $. 
This figure confirms the observation made in the case of real photons in Ref.~\cite{Gervais2012},
namely that the transverse spectral function at $\omega = k$ is enhanced by the 
presence of non-zero $\mu\,$. 
For $\omega > k\,$, it appears that the chemical potential does not dramatically
affect the spectral function compared to the $\mu = 0$ limit.
Above the light cone point, there is a range of energies $k \lsim \omega \lsim 2k$ in which both $\rho_{\rm T}$ 
and $\rho_{\rm L}$ are slightly suppressed, although not as significantly 
as expected on the basis of the free result alone~\cite{Dumitru1993}. 
We note that for large-$\omega$, the OPE predicts an overall enhancement of the rate due to
\begin{widetext}
\bea
 \rho_{\rm V} & \simeq &  \frac{ \Nc M^2 }{4\pi}  
 + 4 g^2 \CF \Nc 
 \biggl\{ 
  \frac{3M^2}{4(4\pi)^3}  
  + \frac{\pi \big( \omega^2 + \frac{k^2}3 \big)
  }{36 M^4} \Big( T^4 + \frac6{\pi^2} T^2 \mu^2 + \frac3{\pi^4} \mu^4 \Big)
 \biggr\}
 \label{OPE V}
 \;, \hspace*{5mm} \\ 
 \rho_{00} & \simeq &  - \frac{ \Nc k^2 }{12\pi} 
 - 4 g^2 \CF \Nc 
 \biggl\{ 
  \frac{k^2}{4(4\pi)^3}  
  + \frac{\pi k^2
  }{108 M^4} \Big( T^4 + \frac6{\pi^2} T^2 \mu^2 + \frac3{\pi^4} \mu^4 \Big)
 \biggr\}
 \;. 
  \label{OPE 00}
\eea
\end{widetext}
The neglected terms are of order 
${\cal O}\big( \frac{T^6}{M^4}, \frac{T^4 \mu^2}{M^4}, \frac{T^2\mu^4}{M^4}, \frac{\mu^6}{M^4}\big)\,$ and beyond.
Therefore, in the large $M^2$ limit  both the transverse and longitudinal rates will
 be {\em enhanced} by the presence of a chemical potential, 
which is just starting to be visible for $\omega \gsim 15\, T$ in Fig.~\ref{fig:spf} (although the effect is very small).

\subsection{Setting physical units}

From \eq{1}, we may calculate the fully differential production 
rate as a function of $\omega$ and $\bm k$
by integrating over the direction of either $\bm p_+$ or $\bm p_-\,$.
Assuming $\Nf=3\,$, with $Q_u = \frac23$ and $Q_{d,s} = - \frac13\,$,
the rate becomes
\bea
\frac{\dd \Gamma_{\ell \bar \ell}}{\dd \omega \, \dd^3 \bm k} 
  &=& \frac{ 2\, \alpha_{\rm em}^2 \, \nB(\omega) 
  }{9\pi^3 \, M^2 } 
\  B \Big( \frac{m_\ell^2}{M^2} \Big) 
\, \rho_{\rm V}(\omega, \bm k) \, ,
\label{M distribution} 
\eea
where we used $K^\mu \rho_{\mu\nu}=0$ to simplify the contraction
with the leptonic tensor.\footnote{%
    The interpolation table of the emission rates is publicly accessible through the dilepton code used in this study~\cite{dileptoncode}.
}

\begin{figure}[!htbp]
  \includegraphics[width= 0.95\linewidth]{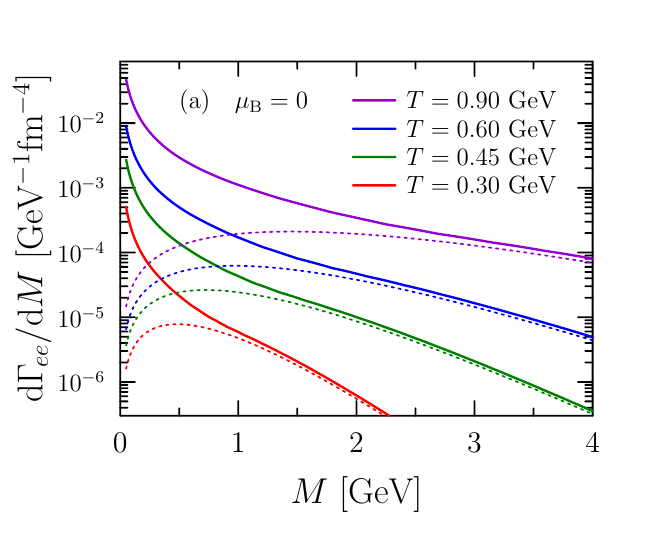}
  \vskip -6mm
  \includegraphics[width= 0.95\linewidth]{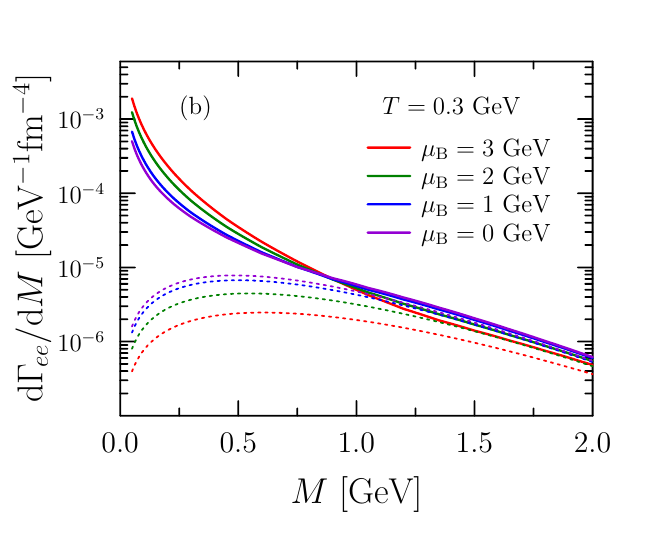}
  \vskip -3mm
  \caption{\label{fig:perturbative rate} 
  Thermal dilepton mass spectrum from a static source, for (a) $\muB=0$ at several temperatures, and (b) for several values of $\muB$ at $T=300$~MeV.
  The solid curves show the NLO rate with a fixed coupling $\alpha_s=0.3$ and the dotted lines depict the rate due to the LO spectral function. In both plots, $\Nf=3$.  
  }
\end{figure}

Throughout this section (so far), 
we have made use of the scaling behaviour of the perturbative result:
$$ 
\frac{\rho_i}{T^2} = \hat \rho_i \Big( \, 
\frac{\omega}{T} \, ,
\frac{k}{T} \, ,
\frac{\muB}{T} \, 
;  \, \alpha_s  \Big) \, ,
$$
where $\hat \rho_i$ is a dimensionless function of dimensionless variables
and we work with natural units ($\hbar = c = k_{_{\rm B}} \equiv 1\,$).
Going over to physical units, we restore
the factor ${\rm GeV}^4{\rm fm}^4 \simeq 1/(0.197327)^4$
when displaying $\dd \Gamma_{\ell \bar \ell}/(\dd \omega \, \dd^3 \bm k)\,$~\cite{Laine:2013vma}.
In addition, the perturbative parameter $\alpha_s$
is fixed in the rest of this work at $\alpha_s = 0.3$, a value consistent with that extracted from work devoted to RHIC phenomenology~\cite{Arslandok:2023utm}.

Let us start with the production rate in the local rest frame of a static thermal source before
generalizing the discussion to an expanding inhomogeneous system.
Presumably, the main features of the invariant mass spectrum
\bea
\frac{\dd \Gamma_{\ell \bar \ell}}{\dd M} &=&
\int_{\bm k} \ \frac{M}{\sqrt{M^2+k^2}} \  \frac{\dd \Gamma_{\ell \bar \ell}}{\dd \omega\, \dd^3 \bm k}
\, ,
\label{dGammadM}
\eea
will carry over to more sophisticated circumstances.

\begin{figure}[t]
    \vskip -4mm
  \includegraphics[scale=.8]{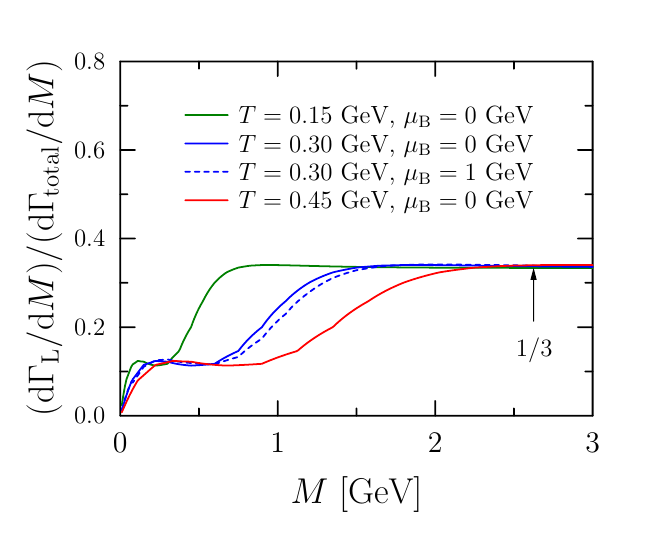}
  \vskip -3mm
  \caption{\label{fig:polarisation rate} 
  Fractional contribution 
  of the longitudinal polarization
  to the total differential rate ($\dd \Gamma_{\rm total} = 2 \dd \Gamma_{\rm T}+\dd \Gamma_{\rm L}\,$) due to a 
  point source Eq.~\eq{dGammadM}.
  Note that for $M\to 0$ only the two transverse degrees of freedom contribute, whereas for large $M$
  the two spectral functions become equal: 
  $\rho_{\rm T} \approx \rho_{\rm L}$.
  }
\end{figure}

The result of inserting \eq{M distribution} into
\eq{dGammadM} is shown in Fig.~\ref{fig:perturbative rate},
which illustrates how the $M$-distribution is affected
by $T$ and $\muB$.
In this figure, the LO results from Eqs.~\eq{rhoV_LO} and \eq{rho00_LO} are also 
shown to illustrate the relative importance of QCD corrections.
We note the role $T$ plays in the absolute magnitude of the rate,
as expected for dimensional reasons.
However, the impact of $\muB$ is more subtle: There 
is evidently some enhancement at low $M$ when increasing $\muB\,$,
but this becomes a very mild {\em suppression} in the intermediate range.
On the grounds of the LO result alone, one might conclude that adjusting $\muB$
to be different from zero leads to a depletion of quarks or antiquarks;
thus fewer dileptons~\cite{Dumitru1993}. 
This effect is apparently partly mitigated at NLO by the
effects of screening and the LPM effect due to 
larger thermal masses~\eq{thermal masses}. 

Since both projections of the spectral function have been determined,
we can decompose the (total) invariant mass distribution
from \eq{dGammadM}
into its transverse and longitudinal components.
This can be accomplished by simply replacing $\rho_{\rm V}$
with $\rho_{\rm T}$ or $\rho_{\rm L}$ in Eq.~\eq{M distribution}.
The fraction contribution of the longitudinal 
polarization is shown in Fig.~\ref{fig:polarisation rate},
for a thermal source at rest.
Evidently, the two transverse components dominate the rate for low $M$
but for highly virtual photons, $\dd \Gamma_{\rm L} = \dd \Gamma_{\rm T}$
and each contributes $\frac13$ of the total rate.
As Fig.~\ref{fig:polarisation rate} demonstrates,
for temperatures and baryon densities of relevance,
the latter situation is realized for $M \gsim 1$~GeV.
Thus we shall not consider the polarizations separately
in the rest of this paper, 
however it would be interesting to consider if the
distinction could be made experimentally,
perhaps from another observable derived from Eq.~\eq{1}.

\section{Dilepton production at RHIC\label{sec:pheno}}

\subsection{Embedding rates in simulations}\label{sec:simulate_dilepton}

Observable spectra are determined by superimposing
the differential rate in the LRF\footnote{%
Here we assume the virtual photon's production takes place on 
time and distances smaller than those of the hydrodynamic evolution, 
and neglect finite-size effects on the scattering amplitudes.
}
\bea
 \frac{\dd N_{\ell \bar \ell}}{\dd t \, \dd^3 \bm x  \, \dd \omega \, \dd^3 \bm k}
 &=& 
 \frac{\dd \Gamma_{\ell \bar \ell}}{\dd \omega \, \dd^3 \bm k} 
 \, ,
 \label{eq:em_rate}
\eea
on top of an expanding hydrodynamical background.
The absolute yield is calculated from a 
spacetime integration over the QGP,
which takes into account the local variation in its thermodynamic
quantities $T \to T(t,\bm x)$ and $\muB \to \muB(t, \bm x)\,$ 
and the boost of $K_\mu=(\omega, \bm k)$ to $K^\prime_\mu$ by the fluid flow velocity $u^\mu(t, \bm x)$; $K^\prime_\mu$ is then the dilepton four-momentum in the lab frame. More explicitly, the emission rate in the lab frame is given by
\bea
    \frac{\dd N_{\ell \bar \ell}}{\dd^4 K^\prime} 
    &=&
    \int \dd t 
    \int \dd^3 \bm x
    \left.\frac{\dd \Gamma_{\ell \bar \ell}}{\dd \omega \, \dd^3 \bm k}\right|_{ K^\mu=\Lambda^{\mu\nu}K^\prime_\nu}
    \ ,
    \label{eq:emission_lab}
\eea
where $\Lambda^{\mu\nu}$ is the Lorentz transformation used to boost the lab frame four-momentum $K^\prime_\mu$ to $K_\mu$ in the LRF of the fluid cell with four-velocity $u^\mu(t, \bm x)$\footnote{
    In the simulation, the four-velocity vector $u^\mu(t, \bm x)$ obtained from MUSIC is represented in Milne coordinates, while both $K_\mu$ and $K^\prime_\mu$ are expressed in Cartesian coordinates. We perform a conversion of $u^\mu(t, \bm x)$ from Milne coordinates to Cartesian coordinates to determine the boost transformation $\Lambda^{\mu\nu}$.
} 
\cite{Ryblewski:2015hea,Burnier2015}. In practice, the spacetime integration is done by summing up contributions from discretized fluid cells. In this study, our primary focus is on thermal dileptons produced from the QGP. Therefore, we specifically consider the fluid cells with temperatures above the freeze-out line found in Ref.~\cite{Cleymans:2005xv}, as shown in Fig.~\ref{fig:phase_diagram_traj}. 
This line closely corresponds to the chemical freeze-out line as extracted by the STAR Collaboration~\cite{STAR:2017sal}. Consequently, we attribute in this work the thermal dileptons emitted from the fluid cells below this line to contributions from hadronic matter. 

Following the usual convention, we align the beam direction with the $z$-axis
and introduce hyperbolic coordinates for the lab frame:
\bea
K^{\prime\,\mu} &=&
\big(
M_\perp \cosh y  ,\,
\bm k_\perp
,\, M_\perp \sinh y
\big) \, ,
\label{Kprime}
\eea
where $M_\perp \equiv \sqrt{M^2 + {\bm k}_\perp^2}$
is the transverse mass
and $y$ is the dilepton's rapidity (we omit the prime on $\bm k_\perp$, the transverse momentum, for brevity).
Let $\phi$ be the azimuthal angle made by $\bm k_\perp$ w.r.t. the $z$-axis. 
Using 
$d^4K^\prime =MdM \, dy \, k_\perp dk_\perp \, d\phi\,$,
the absolute dilepton yield is given by
\be
    \frac{\dd N_{\ell \bar \ell}}{\dd M\, \dd y}  
    \ =\ 
M\int_{k_{\rm min}}^{k_{\rm max}}
\dd k_\perp \, k_\perp
\int_0^{2\pi} {\rm d}\phi \  
\frac{\dd N_{\ell \bar \ell}}{\dd^4 K^\prime} \, ,
\label{convolution}
\ee
where $k_{\rm min}$ and $k_{\rm max}$ are set by detector acceptance. All the quantities involved in Eq.~\eqref{convolution} are defined in the lab frame. The dilepton spectra within a rapidity window $y_{\rm min}<y<y_{\rm max}$ can be further obtained from
\be
    \frac{\dd N_{\ell \bar \ell}}{\dd M \Delta y}  
    &=&
\frac{1}{\Delta y}\int_{y_{\rm min}}^{y_{\rm max}} 
\!\!\dd y \,
\frac{\dd N_{\ell \bar \ell}}{\dd M \, \dd y} \, ,
\label{eq:spect}
\ee
where $\Delta y=y_{\rm max}-y_{\rm min}$ is the corresponding rapidity bin width\footnote{%
    For the sake of brevity, we shall omit the $\ell \bar \ell$ 
    subscript in $N_{\ell \bar \ell}$ from here on. 
    The left-hand side of Eq. (\ref{eq:spect}) is what is plotted as 
    $\dd N/(\dd M \dd y)$ and $y_{\rm max/min} = \pm 1$ unless noted otherwise.}.

\begin{figure}
    \centering
    \includegraphics[width= .93\linewidth]{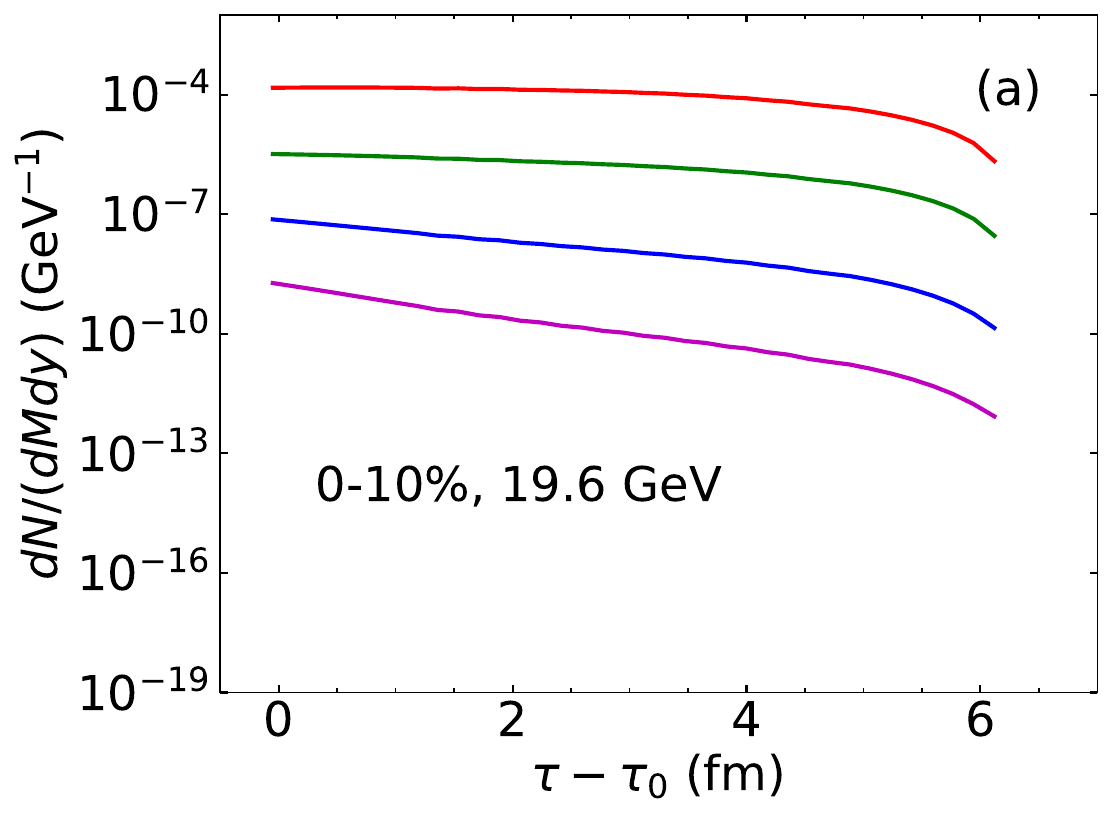}
    \includegraphics[width= .93\linewidth]{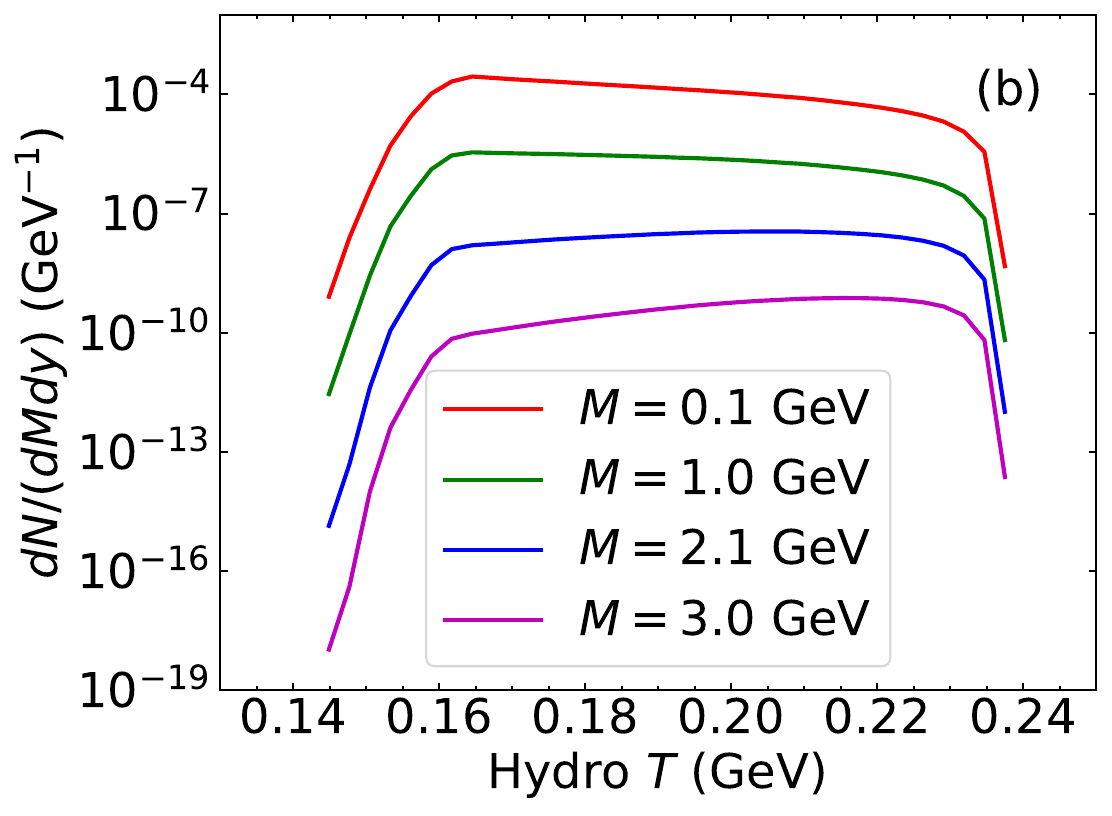}
    \caption{Evolution of dilepton yields with four different invariant masses as functions of (a) proper time  (with $\tau_0$ subtracted) and (b) simulation temperature for 0-10\% Au+Au collisions at $\sqrt{\sNN}=19.6$~GeV.
    }
    \label{fig:dilepton_evolution}
\end{figure}

With the rapid hydrodynamic expansion, the QGP fireball undergoes a dramatic increase in volume and a substantial decrease in temperature. As illustrated in Fig.~\ref{fig:perturbative rate}(a), the emission rate 
$\dd \Gamma_{\ell \bar \ell}/(\dd \omega \, \dd^3 \bm k)$ in Eq.~\eq{eq:emission_lab} exhibits rapid decay as the temperature decreases, 
which stalls dilepton production as the system cools. 
However, simultaneously, the 
spacetime integration $\int d^4 X$ increases with expanding volume, 
and thus bolsters the emission of dileptons. 
These two effects compete, resulting in distinct behaviors of the dilepton yields for various invariant masses emitted from the evolving QGP fireball, as illustrated in Fig.~\ref{fig:dilepton_evolution}.

Figure~\ref{fig:dilepton_evolution}(a) demonstrates that as time elapses, 
dilepton production is suppressed, primarily due to the 
decreasing emission rate resulting from the declining temperature, 
which prevails over the effect of the expanding volume. 
The suppression is also stronger for dileptons with larger invariant masses. 
However, when we organize the fluid cells based on their temperatures, 
calculate the dilepton yields contributed by cells at these temperatures, 
and then plot the dilepton yields as a function of temperature 
in Fig.~\ref{fig:dilepton_evolution}(b), 
we observe distinct behaviors for different invariant masses 
in three temperature ranges. Dileptons originating from the 
high-temperature region ($T\gsim 0.23$~GeV) are primarily emitted by 
fluid cells at the earliest stage of the evolution with the 
highest energy density. 
Conversely, those from the low-temperature region ($T\lsim 0.16$~GeV) 
are radiated from cells near the chemical freeze-out temperature, 
mostly from the latest stage of the evolution. The yields exhibit a significant 
suppression when the temperature decreases to approximately 
$T\approx 0.16\,$GeV. 
This occurs because the fluid cells near midrapidity ($\eta_s \approx 0$), 
which are the primary contributors to dileptons within $|y|<1$, 
cross the freeze-out line (see Fig.~\ref{fig:phase_diagram_traj}). 
Below $T\lsim 0.16$~GeV, most dileptons originate from fluid cells located away from midrapidity, 
characterized by higher chemical potentials and consequently lower freeze-out temperatures. 
Dileptons from these cells with larger spacetime rapidities can still be observed within $|y|<1$, because of 
`thermal smearing,' as discussed in the next section
(and in Appendix~\ref{app:D}).  
As shown in Fig.~\ref{fig:dilepton_evolution}(b), dileptons produced in these two 
regions contribute sub-dominantly compared to those from 
the intermediate temperature range ($0.16$~GeV$\,\lsim T\lsim 0.23$~GeV). 
In this range, we observe an enhancement in dileptons with small masses 
and a suppression in those with large masses as the temperature 
{\em decreases}. 
This suggests that, in certain temperature ranges, 
the expanding volume may dominate over the decreasing 
emission rate for small masses. 
To appreciate Fig.~\ref{fig:dilepton_evolution}(b),
it's important to note that fluid cells with 
lower temperatures predominantly correspond 
to later stages of the evolution, which generally have larger volumes.

\subsection{Dependence on chemical potential and rapidity}\label{sec:muB_and_y}

Figure~\ref{fig:perturbative rate}(b) illustrates the weak $\muB$-dependence of the emission rate. 
It is valuable to investigate the corresponding $\muB$-dependence 
of the dilepton spectra for the baryon-charged QGP, 
as brought up in Sec.~\ref{sec:hydro}. 
This analysis can help to answer whether experimentally measured 
dileptons can effectively serve as a baryometer. 
For this purpose, we compute the dilepton invariant mass 
spectra at the two lowest beam energies (with the highest $\muB$) 
under two distinct scenarios: $i$) a $\muB$-dependent emission rate and 
$ii$) with $\muB$ manually set to zero in the rate. 
Figure~\ref{fig:mu_rapidity_dependence} presents a comparison 
between these two scenarios for 40-50\% Au+Au collisions at 7.7 and 19.6 GeV. 
The analysis demonstrates that the $\muB$-dependence is relatively 
minimal and not resolvable with the current measurement precision. 
This is because even $\muB/T$ achieved at 7.7 GeV 
is relatively small compared to the cases in 
Fig.~\ref{fig:perturbative rate}(b), as shown in Fig.~\ref{fig:phase_diagram_traj}. 
The results illustrate that the total dilepton yields within rapidity window $|y|{\,<\,}1$ 
are not a good baryometer for the systems produced in collisions 
at $\sqrt{\sNN}\geq 7.7$\,GeV.

\begin{figure}[!tb]
    \centering
    \includegraphics[width= .93\linewidth]{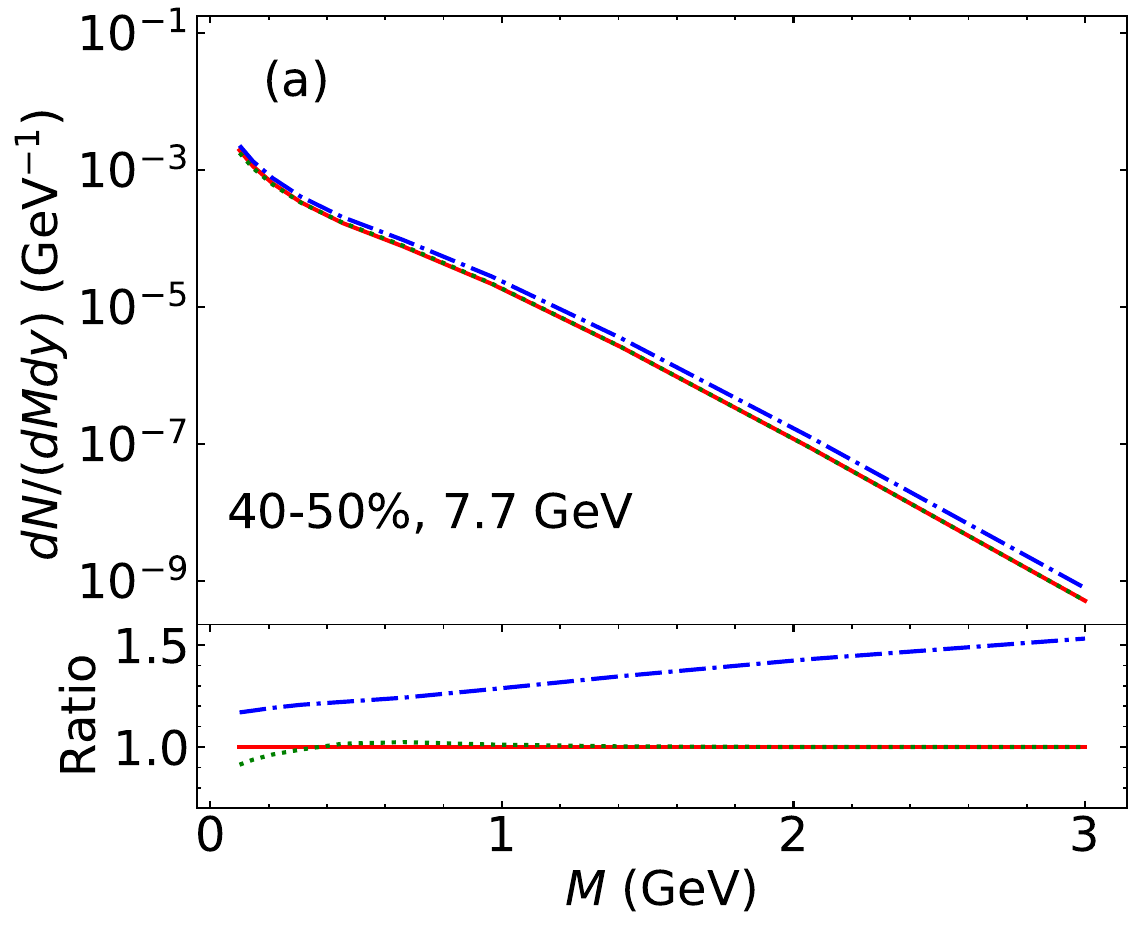}
    \includegraphics[width= .93\linewidth]{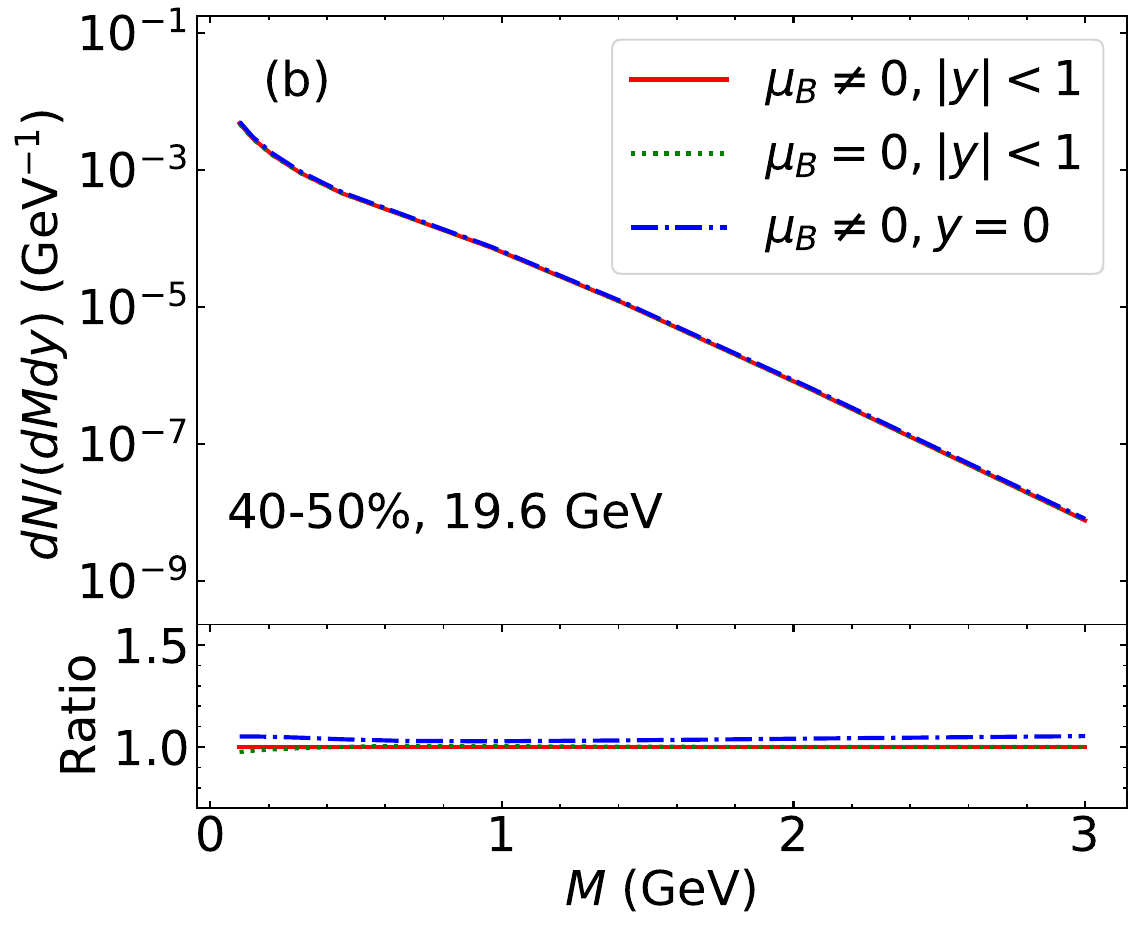}
    \caption{Dilepton invariant mass spectra for 40-50\% Au+Au collisions at 
    (a) $\sqrt{\sNN}=7.7$~GeV and (b) $\sqrt{\sNN}=19.6$~GeV 
    presented in three scenarios: 
    $i$) Spectra within a rapidity window $|y|<1$ with $\muB$-dependent emission rates (red solid line). 
    $ii$) Spectra within a rapidity window $|y|<1$ with $\muB$ manually set to zero in the rate (green dotted line). 
    $iii$) Spectra at midrapidity $y=0$ with $\muB$-dependent emission rates (blue dot-dashed line). The bottom panels display the ratios of the spectra with the first scenario as the denominator.}
    \label{fig:mu_rapidity_dependence}
\end{figure}

Another significant feature of low collision energies, 
in addition to the nonzero baryon chemical potential, 
is the strong violation of Bjorken boost invariance.
This violation is anticipated to be most pronounced 
at the lowest beam energies, manifesting as stronger flow 
and more significant variations in thermodynamic properties 
along the beam direction \cite{Du:2022yok,Du:2023gnv}. 
To investigate the impact of boost-non-invariant effects 
on dilepton production and assess the validity of 
the midrapidity spectra as an approximation for a rapidity window $|y|{\,<\,}1$, 
we have additionally computed the spectra at 
midrapidity $y{\,=\,}0$ while considering $\muB$-dependent emission rates. 
The ratio of the spectrum in this case to 
the one within $|y|{\,<\,}1$ exhibits a significant enhancement at $7.7$~GeV, 
reaching an excess as high as 50\%, and a more modest enhancement at $19.6$~GeV, 
for which the relative excess is $\lsim 10$\%. We have checked that this enhancement 
gradually diminishes at higher beam energies. 
This indicates the significance of boost-non-invariant effects 
on dilepton production at beam energies $\sqrt{\sNN}\lsim10$~GeV. 
Furthermore, it suggests that at beam energies above tens of GeV, 
the spectra at midrapidity $y{\,=\,}0$ serve as a suitable approximation 
for a rapidity window $|y|{\,<\,}1$. 
It's also intriguing to observe that 
the enhancement exhibits a dependence on the invariant mass, 
with a smaller effect for smaller masses. 
This behavior is attributable to the fact that the dilepton 
yield experiences more significant thermal smearing in 
rapidity for smaller masses (see Appendix \ref{app:D}). 
Consequently, the yields at midrapidity $y{\,=\,}0$ 
and within a rapidity window $|y|{\,<\,}1$ begin to coincide for smaller masses.

\subsection{Temperature extraction from dilepton spectra}

The dilepton invariant mass spectrum remains the same under 
change of reference frame, which means it remains unaltered by the collective 
expansion of the medium\footnote{
Strictly speaking, $y$ can still cause $\dd N/ (\dd M \dd y)$ to become modified under a change of reference frame.}.
Consequently, unlike the transverse momentum spectra of photons, 
the dilepton mass spectrum does not experience a blue shift 
and thus is often considered a reliable thermometer of the QCD fireball \cite{Rapp:2014hha,HADES:2019auv}.
Therefore, measurements of real and virtual photons can be combined to constrain both the local and global behaviour of the strongly interacting medium. 

In the limit $M\gg T$ and $\muB$, 
the LRF emission rate can be approximated by 
\begin{equation}
    \frac{\dd \Gamma_{\ell \bar \ell}}{\dd M}\propto (MT)^{3/2}\exp(-M/T)\,.
    \label{eq:nonrel_limit}
\end{equation}
Assuming the spacetime-integrated spectrum 
(see Fig.~\ref{fig:mu_rapidity_dependence}) exhibits a similar form, 
an effective temperature $T_{\rm eff}$ can be extracted from it. 
Importantly, this temperature is termed ``effective'' because 
the evolving fireball is highly dynamic and inhomogeneous and 
the final spectrum contains dileptons that are produced at different 
times and temperatures, as illustrated in Fig.~\ref{fig:dilepton_evolution}. 
With access to the true temperature through hydrodynamic evolution, 
we can  test this extraction method 
and gain insights on the effective temperature by contrasting 
its values obtained from the spectra with the temperatures 
of the fluid cells responsible for these spectra.

In practice, we determine $T_{\rm eff}$ from the 
inverse slope of $\ln[(\dd N/\dd M) M^{-\frac{3}{2}}]\,$.
(The factor $M^{-\frac{3}{2}}$
is included to enhance the linearity with respect to $M$ in 
a log-linear plot.)
We will make use of dileptons with a mass range of $1$~GeV~$<M<3$~GeV, 
which are predominantly generated by the QGP; lower invariant masses will get important contributions from reactions involving composite hadrons \cite{Rapp:2009yu} which are not included in this study. 
In this mass range, Eq.~\eqref{eq:nonrel_limit} can be used as an ansatz
(the suitability of which can be tested by comparing the ``average'' simulation $T$ with $T_{\rm eff}$).

\begin{figure}[!tb]
    \centering
    \includegraphics[width=\linewidth]{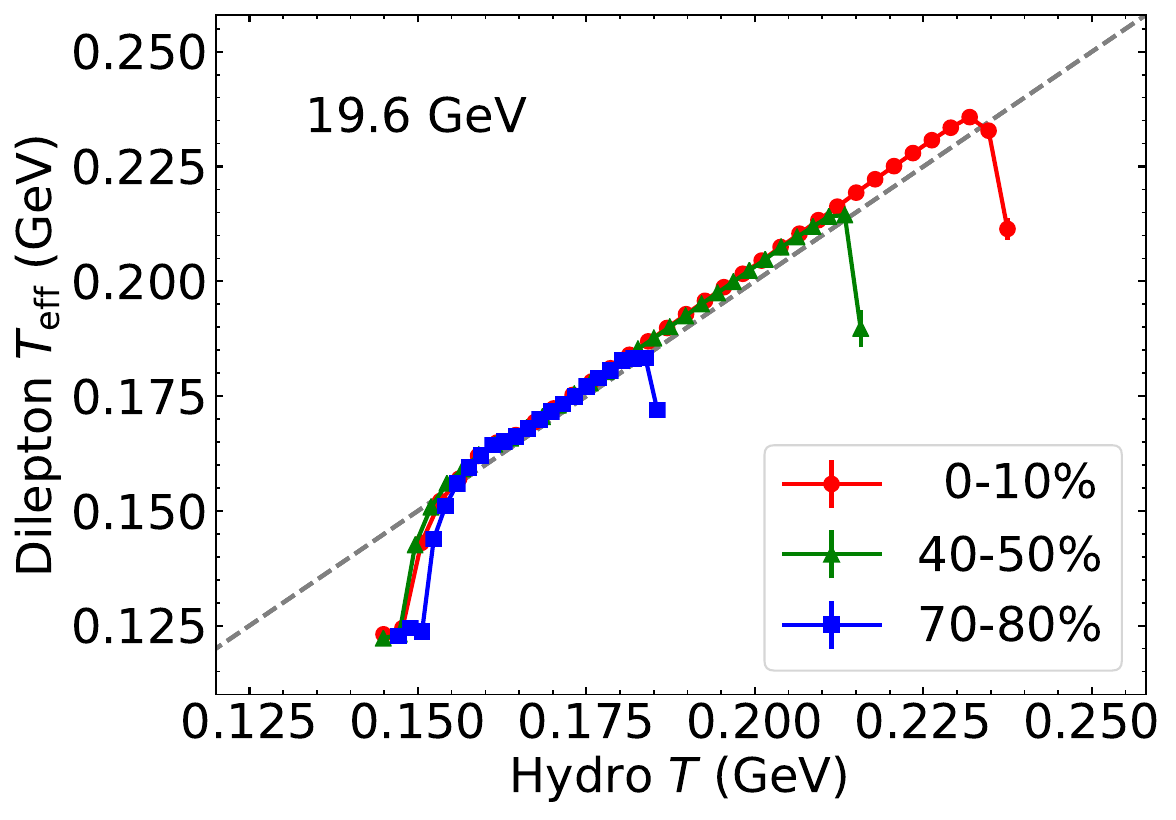}
    \caption{Effective temperature $T_{\rm eff}$ 
    read off dilepton spectrum (``dilepton $T_{\rm eff}$'') 
    emitted from fluid cells at different temperatures (``hydro $T$'') 
    for Au+Au collisions at $\sqrt{\sNN}=19.6$~GeV 
    within three centrality classes. 
    The two temperatures are exactly equal on the grey dashed line.
    }
    \label{fig:temp_extract_T}
\end{figure}

In Fig.~\ref{fig:dilepton_evolution}(b), 
we have computed dilepton spectra produced from 
fluid cells categorized by their respective temperatures. 
Using the temperature extraction method described above, 
we can obtain an effective temperature from the dilepton spectrum 
(``dilepton $T_{\rm eff}$'') corresponding to each fluid temperature 
(``hydro $T$''), and their comparison is shown in 
Fig.~\ref{fig:temp_extract_T}. 
In the temperature range $0.16$~GeV~$\lesssim T\lesssim 0.23$~GeV, 
we note that the two temperatures are remarkably similar, 
suggesting that the temperature extraction method performs well.\footnote{%
    In terms of the physics involved, this may seem somewhat straightforward; 
    however, from a computational perspective, 
    it represents a highly intricate validation of the 
    numerical implementation of our entire framework.
} 
However, 
small deviations can be observed which may arise due to several factors. 
First, Eq.~\eqref{eq:nonrel_limit} used for temperature extraction 
only represents an approximation. 
Second, each ``hydro $T$'' corresponds to a finite temperature range 
(of bin width $\Delta T \simeq 0.002$~GeV) 
rather than a precise value. 
Thirdly, even when the fluid cells share the same temperature, 
they have different proper times in the evolution and thus 
possess different volumes. 
These factors spoil the idealization above and introduce
further complexities into the invariant mass spectra. 
As previously discussed in Sec.~\ref{sec:simulate_dilepton}, 
dileptons emitted from the high-temperature region ($T\gtrsim 0.23$~GeV) 
and the low-temperature region ($T\lesssim 0.16$~GeV) 
originate from fluid cells with the highest energy densities or those 
near the chemical freeze-out temperature, respectively. 
Although the deviations in these regions are slightly more pronounced, 
the contributions from fluid cells within the range 
$0.16$~GeV~$\lesssim T\lesssim$~$0.23$~GeV 
dominate the dilepton production.

\begin{figure}
    \centering
    \includegraphics[width=0.93\linewidth]{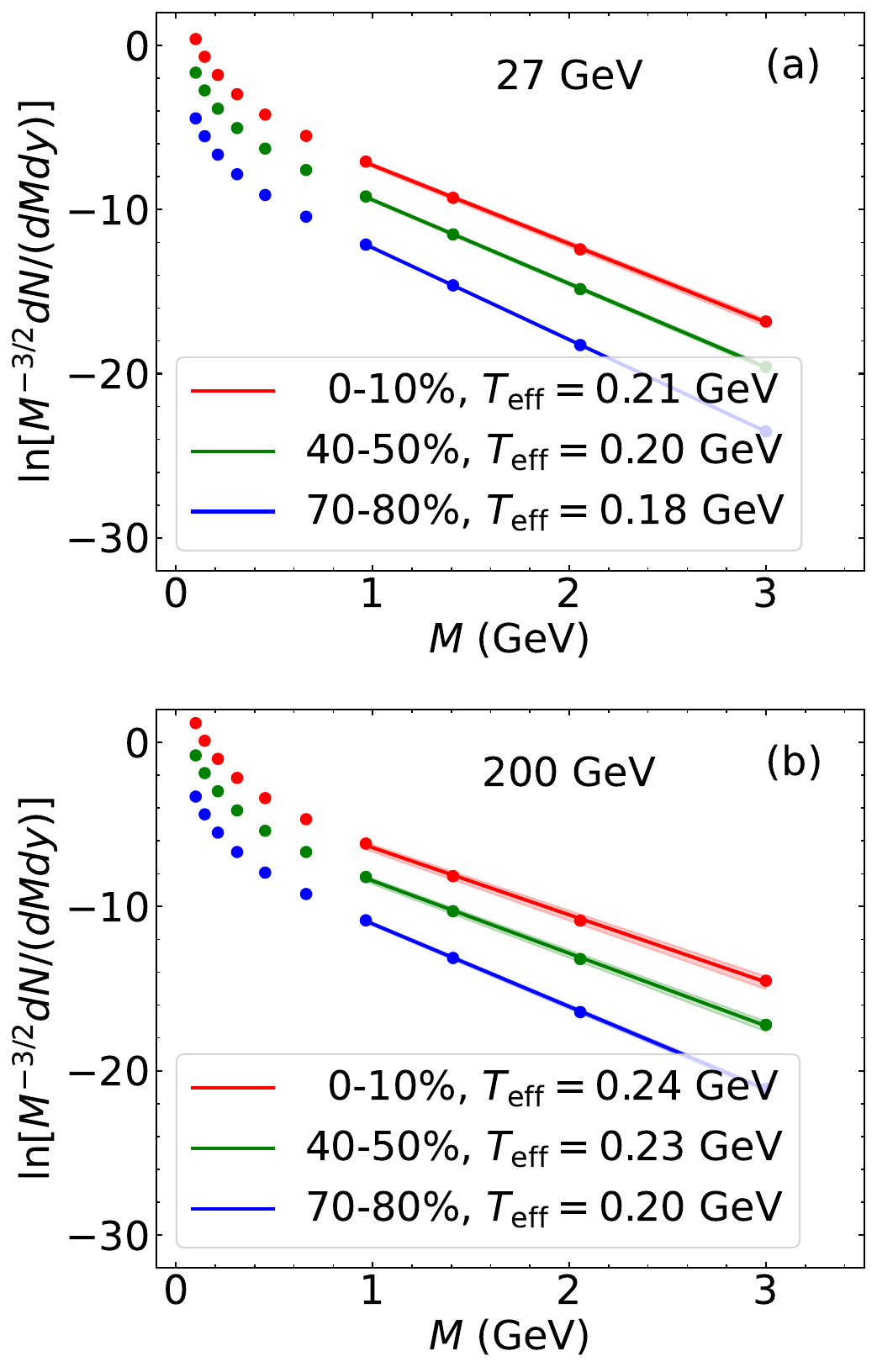}
    \caption{Effective temperature extraction from dilepton spectrum for Au+Au collisions at 
    (a) $\sqrt{\sNN}=27$~GeV and 
    (b) $\sqrt{\sNN}=200$~GeV
    within three centrality classes. The straight lines represent the linear fit function in the mass window 
    $1$~GeV~$\leq M\leq$~$3$~GeV, with their inverse slope providing the effective temperature $T_{\rm eff}$. The shaded band indicates the uncertainties associated with the fitting procedure.}
    \label{fig:temp_extract}
\end{figure}

With the temperature extraction method validated across 
various fluid cell temperatures, we proceed to apply the method to 
the spacetime-integrated dilepton spectra produced 
throughout the entire evolution and determine an effective temperature for the expanding fireball. 
Evidently, the resulting $T_{\rm eff}$ would represent an average over the temperatures of fluid cells above the freeze-out line. 
Figure~\ref{fig:temp_extract} illustrates this implementation, 
using dileptons with $1$~GeV~$\leq M\leq 3$~GeV for temperature extraction. 
In the plot, the dots correspond to the model calculations, 
and the lines represent the linear fit functions. 
The inverse slope of the fit function provides the value of $T_{\rm eff}$. 
The shaded band in the figure represents the uncertainties resulting from 
the fitting procedure, consequently leading to 
uncertainties in $T_{\rm eff}$. 
It's noteworthy that the linear fit performs exceptionally well 
within the log-linear plot, strongly indicating that 
the approximate expression in Eq.~\eq{eq:nonrel_limit} is indeed a 
reasonable approximation, 
with the exponential term $\exp(-M/T)$ playing a dominant role 
in the mass window $1$~GeV~$\leq M\leq 3$~GeV.

\begin{figure}
    \centering
    \includegraphics[width= .95\linewidth]{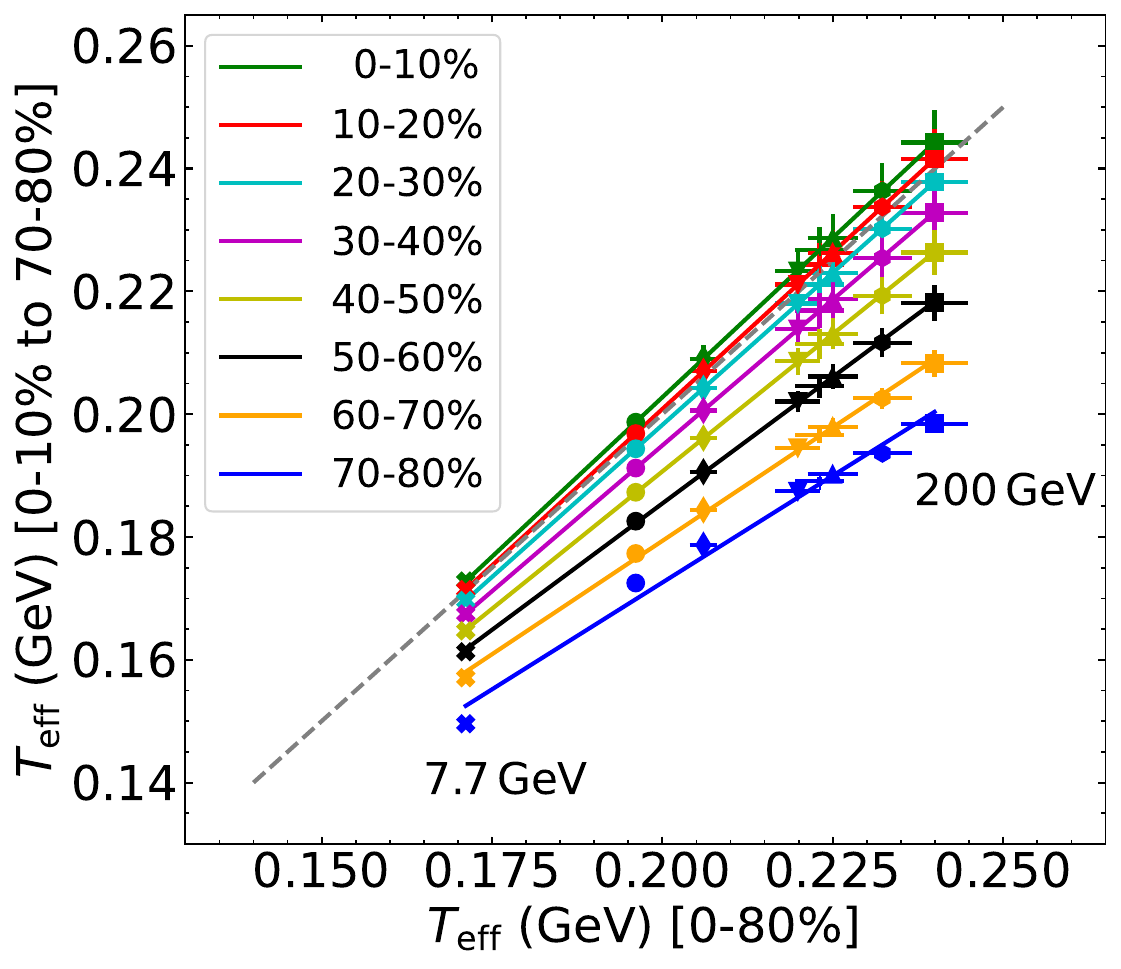}
    \caption{
        Effective temperature $T_{\rm eff}$ derived from dilepton spectra for Au+Au collisions at eight beam energies from 
        $\sqrt{\sNN}=7.7$~GeV (leftmost markers) to $200$~GeV (rightmost markers). 
        Horizontal coordinates of the markers represent $T_{\rm eff}$ from the minimum-bias collisions (0-80\%), while vertical coordinates show $T_{\rm eff}$ within smaller centrality bins (0-10\%, 10-20\%, \ldots, 70-80\%). 
        Error bars denote associated uncertainties, and straight lines represent linear fits applied to the markers.
    }
    \label{fig:temp_centralities}
\end{figure}

Using the extraction method, 
an effective temperature can be derived from a dilepton 
spectrum of a system at any beam energy within a given centrality class. 
We calculate the dilepton spectra for Au+Au collisions at the eight beam energies studied in this work, considering centrality bins ranging from 0-10\% to 70-80\%, as well as the minimum-bias 0-80\% centrality class which aligns with the experimentally measured centrality by STAR \cite{STAR:2013pwb,STAR2015,STAR:2015zal,STAR:2023wta}. 
The agreement between our calculated spectra and experimental data in the minimum-bias events is discussed in the companion Letter \cite{Churchill:2023zkk}. 
Effective temperatures derived from these spectra are shown in Fig.~\ref{fig:temp_centralities}. 
In the plot, the horizontal coordinates of the markers illustrate $T_{\rm eff}$ values obtained from the minimum-bias collisions, while the vertical coordinates display $T_{\rm eff}$ values within narrower centrality bins. 

The figure shows that, within a specific centrality bin, 
$T_{\rm eff}$ decreases at lower beam energies, while at a given beam energy, 
$T_{\rm eff}$ decreases as one moves from central to peripheral collisions. 
Naturally, these observations are in line with the expectation, 
as they simply reflect the characteristics of the hydrodynamic evolution 
constrained by the hadron yields, 
which was previously discussed in Sec.~\ref{sec:hydro}. 
It's worth highlighting the intriguing observation that the $T_{\rm eff}$ values obtained from minimum-bias collisions at varying beam energies fall between the 10-20\% and 20-30\% centrality collisions, with a closer alignment to the 10-20\% values. This is indeed expected, as the dilepton production in minimum-bias collisions is dominated by central collisions. 
Verifying this observation in future experimental measurements, especially when statistical precision is sufficient, could serve as a critical test of the physics incorporated in this study. 

Figure~\ref{fig:temp_centralities} illustrates a consistent trend: 
uncertainties in $T_{\rm eff}$ rise with increasing beam energy within a specific centrality class, 
or from peripheral to central collisions at a given beam energy. This trend can also be seen in the results presented for the two specific beam energies shown in Fig.~\ref{fig:temp_extract}. 
One contributing factor to this trend is the prolonged lifetime of the fireball at higher beam energies or central collisions, coupled with more significant temperature variations during its evolution. 
Consequently, the dilepton spectra exhibit more pronounced deviations from those associated with a particular effective temperature. 
This phenomenon is likely the main source of the increased uncertainties in $T_{\rm eff}$ extraction.

\subsection{Integrated yields and fireball lifetime}

\begin{figure}[!tb]
    \centering
    \includegraphics[width= 0.93\linewidth]{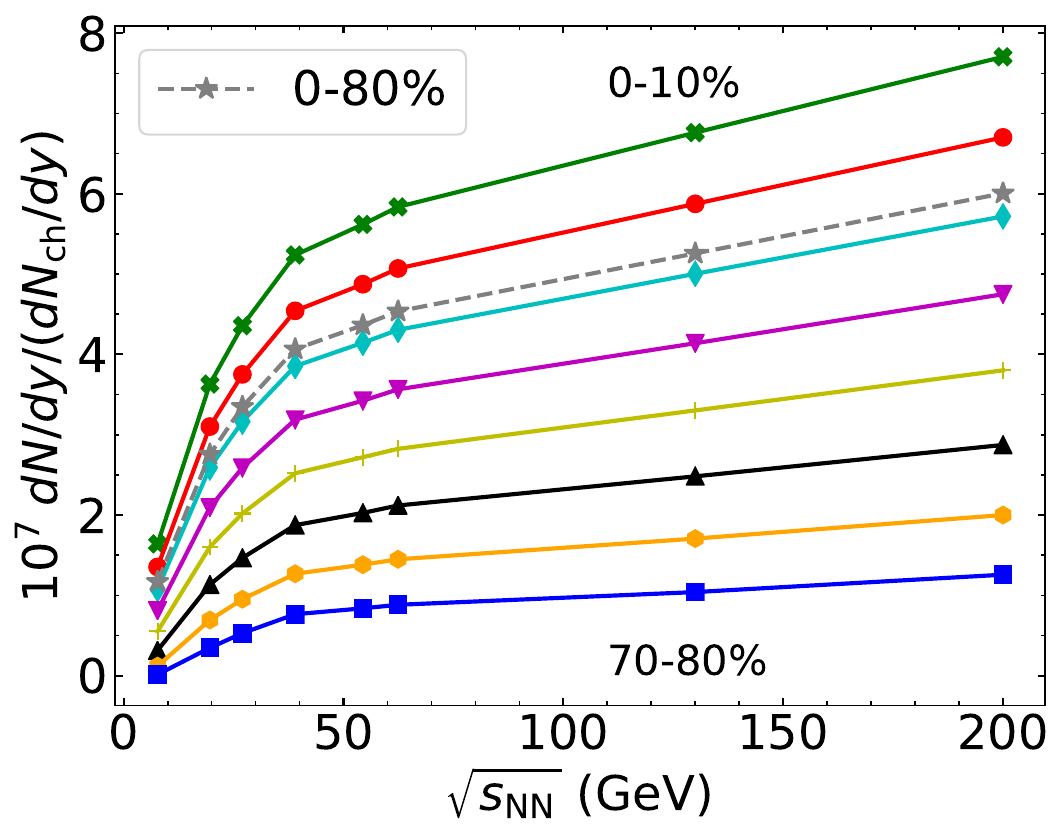}
    \caption{
        Integrated dilepton yields in the mass range 
        $1$~GeV $\leq M \leq 3$~GeV, 
        normalized by midrapidity charged hadron multiplicity 
        ($\dd N_{\rm ch}/ \dd y$), 
        for Au+Au collisions at eight beam energies spanning from 
        $\sqrt{\sNN}=7.7$~GeV (leftmost markers) to 
        $200$~GeV (rightmost markers). Results are presented within centrality classes ranging from 
        0-10\% (uppermost curve) to 70-80\% (lowermost curve), 
        in 10\% increments. The dashed line  represents results within centrality class 0-80\%.
        }
    \label{fig:integrated_yields}
\end{figure}

The dilepton yields in the low-mass region ($M \lesssim 1$  GeV) are thought to correlate with the total lifetime of the nuclear fireball~\cite{Rapp:2014hha,STAR:2015zal}. 
In that region, dilepton production is dominated by hadronic sources 
which are not within the scope of this study. 
Therefore, we investigate the relationship between thermal yields in the intermediate mass region, dominated by QGP thermal radiation, 
and the duration of the QGP stage as determined by our realistic hydrodynamic modeling. 

In Fig.~\ref{fig:integrated_yields}, we present integrated dilepton yields within the mass range of $1$~GeV~$\leq M \leq 3$~GeV, 
normalized by midrapidity charged hadron multiplicity ($\dd N_{\rm ch}/\dd y$). 
We estimate $\dd N_{\rm ch}/\dd y$ by the $\dd N/\dd y$ sum of $\pi^\pm,\,K^\pm,\,p$ 
and $\bar p$ obtained from the model calculation, 
following the STAR measurements. The figure illustrates a clear trend: 
within a specific centrality class, the yield decreases at lower beam energies, and does so more rapidly in central than in peripheral collisions. 
Similarly, at a given beam energy, 
the yield decreases from central to peripheral collisions. 
These trends are consistent with our expectations regarding the QGP stage's lifetime. 
The results for the 0-80\% centrality class are also displayed, and they fall between the results for the 10-20\% and 20-30\% centrality classes, with a closer alignment to the latter. 
This observation is similar to what we've noticed regarding the effective temperatures shown in Fig.~\ref{fig:temp_centralities}, with the exception that the effective temperatures of the 0-80\% centrality appear closer to those of the 10-20\% centrality.

The QGP stage's lifetime, denoted as $\tau_{\rm QGP}$, 
is calculated as the difference between $\tau_{\rm fo}$ 
(the time when the temperatures of all fluid cells within $|\eta_s|<1$ reach the chemical freeze-out line) 
and $\tau_0$ (the starting time of the hydrodynamic description). 
This corresponds to the time range covered by the fluid cells contributing to the thermal dilepton yields, 
defined as the difference between the maximum and minimum times 
associated with all fluid cells. 
The correlation between normalized integrated dilepton yields 
and QGP lifetimes is depicted in Fig.~\ref{fig:lifetime_vs_yields}. 
In this plot, results within distinct centrality classes at 
a given beam energy are connected by lines for clarity. 
The figure illustrates a roughly linear relationship between normalized integrated dilepton yields and QGP lifetimes at specific beam energies $\sqrt{\sNN}\geq 19.6$~GeV; 
however, it's important to note that this proportionality is not consistent across all beam energies.

\begin{figure}[!tb]
    \centering
    \includegraphics[width=0.93\linewidth]{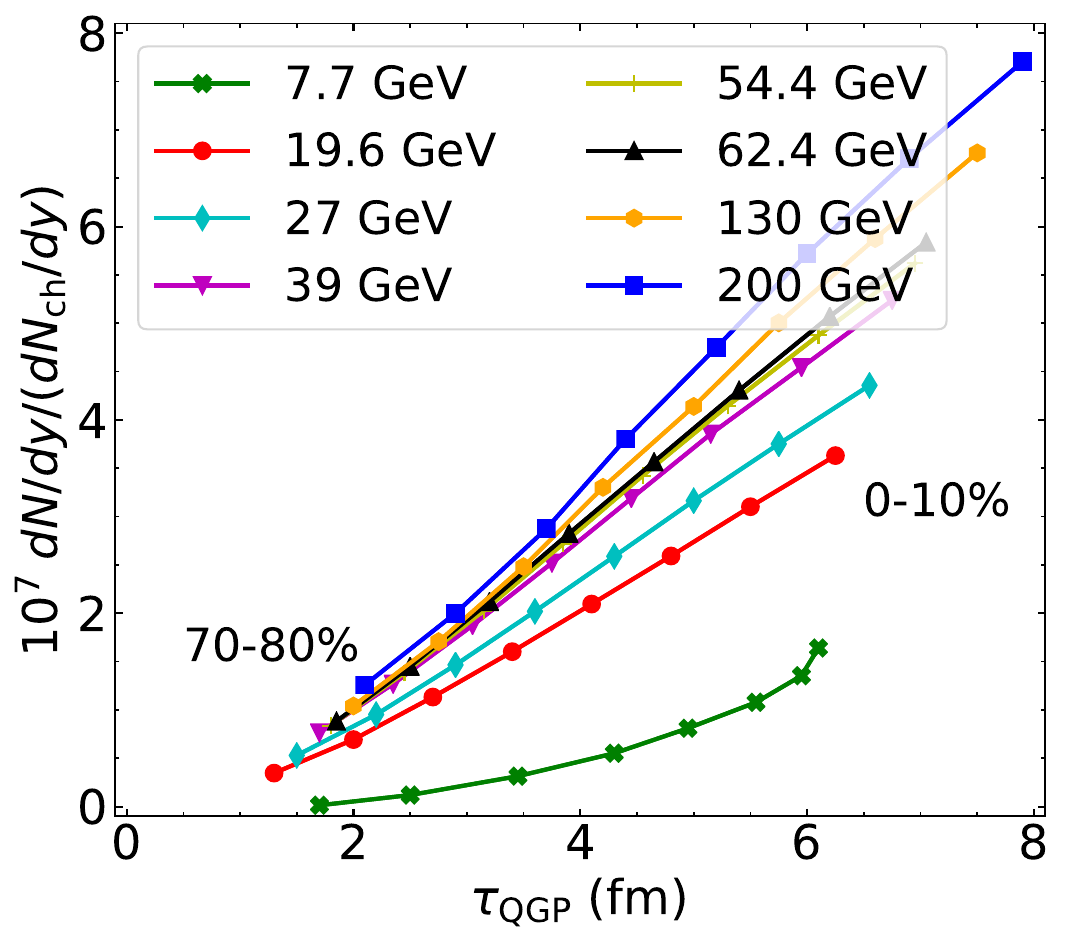}
    \caption{
        Correlation between normalized integrated dilepton yields and QGP lifetimes in Au+Au collisions at eight beam energies, 
        ranging from 
        $\sqrt{\sNN}=7.7$~GeV (lowest curve) to $200$~GeV (highest curve), across centrality classes from 0-10\% (rightmost markers) to 70-80\% (leftmost markers).
    }
    \label{fig:lifetime_vs_yields}
\end{figure}

At $\sqrt{\sNN}=7.7$~GeV, a notably different trend is manifest in Fig.~\ref{fig:lifetime_vs_yields}. 
Firstly, unlike the rough linearity observed at higher energies, 
the correlation at $7.7$~GeV is more complicated.
Secondly, within specific centrality classes, 
we notice a systematic decrease in the QGP stage's lifetime with decreasing beam energy, which remains consistent until $19.6$~GeV. 
At $7.7$~GeV, the system features a high chemical potential corresponding to a low chemical freeze-out temperature 
(see Fig.~\ref{fig:phase_diagram_traj}). This, combined with the notably slower expansion at lower beam energies, leads to a prolonged duration required to reach the freeze-out line and thus a longer QGP  lifetime. 
However, it is important to emphasize that defining the QGP  lifetime becomes challenging at such low beam energies because of the long time needed for the two colliding nuclei to interpenetrate each other. 
In principle, at these low energies a dynamical initialization process would be necessary, which goes beyond the scope of the present study.

\section{Summary and Conclusion}

We have derived rates for thermal dilepton production at NLO in the QCD coupling constant, 
for partonic matter at finite temperature and finite baryon density. 
The establishment of those rates necessitated considering 
the contribution of the two-loop photon self-energy diagrams, 
combined with the re-summed class of diagrams that represent the LPM effect. The new spectral functions obtained from the two-loop topologies are Eqs.~(\ref{masters, mn}) and (\ref{masters, 00}). 
When combining strict one and two-loop results with the LPM effect, 
first advocated by Ref.~\cite{Ghisoiu2014}, 
close attention was paid to double-counting issues, and a spectral function which is finite and continuous across the light cone was obtained. 
The thermal rates for lepton pair production exhibit a large sensitivity to NLO corrections which translates into a large enhancement over the LO result at low invariant masses. 
These effects still persist in the intermediate mass domain, depending on the probed temperature. 
The inclusion  of a  baryon chemical potential boosts the dilepton rates at low masses. 
For intermediate masses, the rates with larger $\muB$ are suppressed with respect to those with smaller values, 
but to a lesser extent than what is found at LO.\footnote{%
In the large $M$ limit, the OPE results, Eqs.~(\ref{OPE V}) and (\ref{OPE 00}), predict an enhancement of the dilepton thermal rates, which however remains modest for the invariant masses and temperatures considered in this work.} 
We have also highlighted the marked difference 
between the transverse and longitudinal components of the dilepton emission rates, in Fig.~\ref{fig:polarisation rate}. The phenomenology resulting from this  will be explored in detail in future work, as will be the effects on the rates of of viscosity corrections.

The second part of our paper is devoted to going from rates to yields, 
using a modern fluid-dynamical approach which has been calibrated to reproduce hadronic observables. 
The temperatures extracted from the slope of the dilepton spectrum for invariant masses $M$ such that 1~GeV$<M<3$~GeV, 
have been compared with those in the hydrodynamic simulation for Au+Au collisions at different energies and centralities (Fig.~\ref{fig:temp_extract_T}). 
This investigation with (3+1)-dimensional viscous fluid dynamics finds that measurements of thermal lepton pairs offer a quantitative assessment of the temperature (Figs.~\ref{fig:temp_extract} and \ref{fig:temp_centralities}) of the QCD medium \cite{Churchill:2023zkk} as well as its lifetime (Figs.~\ref{fig:integrated_yields} and \ref{fig:lifetime_vs_yields}). 
Its baryonic content is more difficult to extract from dileptons, but future measurements with the capacity to extract polarization signature could provide further information. 

In addition to its ability to provide temperature information, 
the intermediate invariant mass domain has been identified as a promising  region 
in which to look for signals of chiral symmetry restoration\footnote{%
The NA60+ Collaboration estimates that chiral symmetry restoration at SPS energies would lead to a difference in the dilepton spectrum of 20-25\%, 
in the window of invariant mass $[0.9, 1.4]$~GeV~\cite{NA60:2022sze}.} \cite{Hohler:2013eba}.
These studies  make it clear that 
a precise characterization of the strongly interacting medium via electromagnetic radiation 
requires state-of-the-art emissivities,
integrated with sophisticated modeling.

\acknowledgments
We are happy to acknowledge very useful discussions with Bailey Forster, Han Gao, and Jean-Fran\c{c}ois Paquet. This work was funded in part by the U. S. Department of Energy (DOE), under Grant No. DE-FG02-00ER41132 (G.~J.), in part by the Agence Nationale de la Recherche,  under Grant No. ANR-22-CE31-0018 (AUTOTHERM) (G.~J.),  and in part by the Natural Sciences and Engineering Research Council of Canada (J.~C., L.~D., C.~G., S.~J.). Computations were made on the B\'eluga supercomputer system from McGill University, managed by
Calcul Qu\'ebec and Digital Research
Alliance of Canada.
\appendix

\renewcommand{\thefigure}{A\arabic{figure}}
\setcounter{figure}{0} 

\section{Master integrals\label{app:A}}

In Ref.~\cite{Jackson2021}, 
the types
of interactions that would contribute to NLO rates were studied
(with full generality) and a numerical 
routine for {\em any} combination of particles, masses, chemical potentials and
 a wide class of matrix elements was developed.
For the dilepton rate, it is preferable to use a more tailored approach
which requires a two-dimensional phase space integration~\cite{Jackson2019a}.
This applies, specifically, to the master integrals introduced
in Eq.~\eq{I def}
for which the code can be found at Ref.~\cite{integralcode}.

To account for non-zero $\mu$, 
the abbreviations defined in Eq.~(6.3) of Ref.~\cite{Jackson2019a}
(where $n_s(\omega)\equiv [\exp(\omega/T)-s]^{-1}$) should
be generalised to the following:
\be
n_0 &=\ s_0 n_{s_0}(\omega)  \qquad  &  , \label{ni} \\
n_1 &=\ s_1 n_{s_1}(p-\mu_1)    &  , \nonu \\
n_2 &=\ s_2 n_{s_2}(q-\mu_2)    &  , \nonu \\
n_3 &=\ s_3 n_{s_3}(r-\mu_3)    &  ; \quad r=\omega-p-q  \, ,\nonu \\
n_4 &=\ s_4 n_{s_4}(\ell-\mu_4) &  ; \quad \ell=\omega-p \, , \nonu \\
n_5 &=\ s_5 n_{s_5}(v-\mu_5)    &  ; \quad v=\omega-q \, . \nonu
\ee
If the particles are in chemical equilibrium, the potentials 
are constrained by linear relations:
 $\mu_3 = -\mu_1 -\mu_2$, $\mu_4 = -\mu_1$ and $\mu_5 = - \mu_2\,$.

Of relevance to dilepton production, 
particles `1' and `4' are quarks with $\mu_1=\mu_4 = \mu$,
particles `2' and `5' are anti-quarks with $\mu_2 = \mu_5 = - \mu$ and
particle `3' is a gluon ($\mu_3 = 0$).
The statistical factors are thus $s_0 = s_3 = +1$ and $s_1 = s_2 = s_4 = s_5 = -1\,$.
For our purpose, we specify \eq{ni} with the replacements,
\bea
& & n_0 \ \to\ +\,\nB(\omega) \, , \hspace{0.9cm}
n_1 \ \to\ -\,\nF(p-\mu) \, , \quad \nonu\\
& & n_2 \ \to\ -\,\nF(q+\mu) \, ,\quad
n_3 \ \to\ +\,\nB(r) \, , \quad \nonu\\
& & n_4 \ \to\ -\,\nF(\ell+\mu) \, , \quad
n_5 \ \to\ -\,\nF(v-\mu) \, . \quad
\eea
The master integrals 
defined in \eq{I def} can be studied independently within the HTL approximation, and we have checked our result in this limit~\cite{Jackson2022}.

\section{Matching of IR-singularities near the light cone\label{app:B}}

For \eq{resummation} to make sense, we expect that the (subtracted) singularity in \eq{LPM T}
is compensated by the strict two-loop result in \eq{masters, mn} and \eq{masters, 00} to render
the functions non-singular.

We focus on $\rho_{\rm V}\,$, but the same can be done for $\rho_{00}\,$.
In \eq{masters, mn}, the discontinuous terms are $\rho_{11010}^{(0,0)}$ and 
$\rho_{1111(-1)}^{(0,0)}\,$ (the latter also being responsible for the log divergence).
Following Ref.~\cite{Jackson2019a}, we express
$\rho_{1111(-1)}^{(0,0)}=
\rho_{10110}^{(0,0)} + K^2 \rho_{11110}^{(0,0)} - \rho_{11110}^\text{\,\large $\star$}\,$.
Therefore,
since $\rho_{11110}^{(0,0)}$ is continuous at $\omega=k\,$,
\bea
\left.\rho_{\rm V}\right|_{\rm disc}  &=& 8 g^2 \CF \Nc \nonu \\
 & \times &\Bigl\{  
  \lim_{\omega^{ }\to k^+_{ }} - \lim_{\omega^{ }\to k^-_{ }}
 \Bigr\} \Big[ 
\rho_{10110}^{(0,0)}
-\rho_{11010}^{(0,0)} 
- \rho_{11110}^\text{\,\large $\star$}
 \Big] \, . \label{disc}\nonu \\
\eea
The first two master integrals above factorise:
\bea
 \rho_{10110}^{(0,0)} &-&\rho_{11010}^{(0,0)} =\nonu \\
 &&\Im\,  \Big\{
\bigg( \sumint{P}\,\frac1{P^2 (K-P)^2} \bigg)
\bigg( \sumint{R}\,\frac1{R^2} - \sumint{R} \,\frac1{Q^2} \bigg) \Big\},\nonu \\
\eea
where we changed integration variables from $Q$ to $R=K-P-Q$ in the second term.
The second factor is real, and proportional to $m_\infty^2$ because
$$
\sumint{R} \, \frac1{R^2} \ = \ \frac{T^2}{12} \, , \qquad
\sumint{Q} \, \frac1{Q^2} \ = \ - \, \frac1{24} \Big( T^2 + 3 \frac{\mu^2}{\pi^2} \Big)  \, .
$$
The first factor is just a one-loop integral 
(the same is encountered in the LO calculation),
it reads
\bean
\Im \, 
\bigg( \sumint{P}\,\frac1{P^2 (K-P)^2} \bigg) &=&
\frac{\omega}{k} \langle 1 \rangle
\eean
which gives a discontinuity
\bea
& & \hspace{-1cm} \Big\{ \rho_{10110}^{(0,0)} -\rho_{11010}^{(0,0)}\Big\} \Big|_{\rm disc} \ = \ 
\frac{ (T^2 + \frac{\mu^2}{\pi^2} ) }{128 \pi k} 
\nonu\\
&\times&
 \int_{-\infty}^\infty \!
 {\rm d}\epsilon 
 \, 
 \Big\{  
 1- \nF^{ }(\epsilon-\mu) - \nF^{ }(k-\epsilon+\mu) 
 \Big\}
 \;. \label{rho tad}
\eea
Where the Cauchy principal value of the integral above is implied.
As for the last master integral in \eq{disc}, 
it has the following discontinuity:
\bea
\left.\rho_{11110}^\text{\,\large $\star$}\right|_{\rm disc}
&=&
\frac{ (T^2 + \frac{\mu^2}{\pi^2} ) }{256 \pi} 
 \int_{-\infty}^\infty \!
 {\rm d}\epsilon 
 \, 
 \bigg\{  \label{rho star}\\
& & \hspace{-.9cm} \big[1- \nF^{ }(\epsilon-\mu) - \nF^{ }(k-\epsilon+\mu) \big] 
 \bigg( \frac{1}{\epsilon}+\frac1{\omega-\epsilon} \bigg)
 \bigg\} \, , \nonu
\eea
which may be obtained after reinstating $\mu$ in Eq.~(5.15) from Ref.~\cite{Jackson2019a}.

Altogether, the discontinuity predicted from \eq{disc} turns out to agree with
Eq.~\eq{LPM T}.
The transverse LPM spectral function has 
\bea
 \rho_{\rm T}^{ } \big|^{(g^2)}_{\rm disc}
 \; = \; 
 \frac{\Nc m_\infty^2}{8\pi}
 \int_{-\infty}^\infty \!
 {\rm d}\epsilon 
 \, 
 \bigg\{   \label{disc_prediction}\\
 & & \hspace{-4.3cm} \big[1- \nF^{ }(\epsilon-\mu) - \nF^{ }(k-\epsilon+\mu) \big]
 \bigg( \frac{2}{k} - \frac{1}{\epsilon} - \frac1{k - \epsilon} \bigg) \bigg\} 
 \;. \nonu
\eea
The $\frac2k$-term in parentheses 
matches with \eq{rho tad}, and the 
$(\frac1{\epsilon} + \frac1{k-\epsilon})$-term
matches with \eq{rho star}.
An explicit numerical demonstration that 
the resummed spectral function is both finite and continuous
across the light cone
can be found in Fig.~(3) of Ref.~\cite{Jackson2022}.

\section{Dilepton rapidity distribution \label{app:D}}
In a simplified system, 
we illustrate the thermal smearing effect 
on the rapidity-dependent yields discussed in Sec.~\ref{sec:muB_and_y}. 
According to Eqs.~(\ref{eq:em_rate}-\ref{convolution}), 
the differential rate from a specific spacetime coordinate
is given by
\bea
\frac{\dd \Gamma_{\ell \bar \ell}}{\dd M \, \dd y}  &=& M
\int \dd^2 \bm k_\perp
\left.\frac{\dd \Gamma_{\ell \bar \ell}}{\dd \omega \, \dd^3 \bm k}\right|_{K^\mu=\Lambda^{\mu\nu}K^\prime_\nu} \, ,
\label{convolution-bjorken}
\eea
where $K^\prime_\mu$ is measured in the laboratory frame
(and thus so are $y$ and $\bm k_\perp$, cf.~Eq.~\eq{Kprime}), 
while $K_\mu$ is the four-momentum in the LRF where the 
integrand is calculated. 
For a volume element flowing with a 
four-velocity $u_\mu(t,\bm x)$, the transformation $\Lambda(t,\bm x)$ from the laboratory frame to the 
LRF leads to
\bea
\omega = K^\prime_\mu u^\mu \, , \quad 
k =  \sqrt{ (K^\prime_\mu u^\mu)^2 - M^2 } \, ,
\label{omega and k}
\eea
for the arguments of the spectral function.

In a boost invariant scenario, 
with Bjorken expansion (in the $z$-direction, i.e. $v_z = z/t$), 
the fluid velocity is assumed to take the form
\bea
u^\mu = \gamma_\perp
\big( \cosh \eta_s, \, \vec{v}_\perp, \, \sinh \eta_s \big)
\, , \quad \gamma_\perp \equiv \frac1{\sqrt{1-\vec{v}_\perp^2}} \, ,
\nonu
\eea
where $\vec{v}_\perp$ is the transverse flow~\cite{Bjorken:1982qr}. 
Hence the replacement \eq{omega and k} reads:
\bea
\omega &=&
\gamma_\perp \big[ M_\perp \cosh(y-\eta_s) - k_\perp v_\perp \cos \varphi \big] 
\, , \nonu\\
k &=& \sqrt{\omega^2 - M^2} \, ,
\label{omega and k boost inv}
\eea
where we used Eq.~\eq{Kprime} and introduced $\varphi$ as 
the azimuth, taken as the angle between $\bm k_\perp$ and $\bm v_\perp\,$.

\begin{figure}[!tbp]
  \centering
  \includegraphics[scale=0.8]{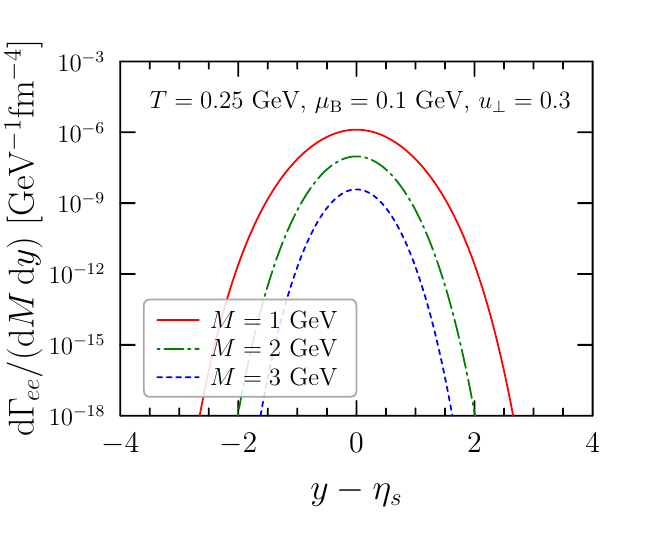}
  \vskip -3mm
  \caption{
     Double differential dilepton rate from Eq.~\eq{convolution-bjorken}, as a function of $y-\eta_s\,$, 
     assuming longitudinal boost invariance.
     The curves are shown 
     for $M = \{ 1,2,3 \}$~GeV
     with $T=0.25$~GeV and $\mu_B=0.1$~GeV and $u_\perp = 0.3\,$.   
}
  \label{fig:yields_in_rapidity} 
\end{figure}

It may be observed that Eq.~\eqref{convolution-bjorken} is peaked around $y\approx\eta_s\,$, because 
the factor $\nB(\omega)$ is largest when $\omega$ is minimized. 
For $M_\perp \gg T$, which is expected to hold for intermediate mass dileptons, 
a source at spacetime rapidity $\eta_s$ mostly emits dileptons with
the same rapidity in momentum-space.\footnote{\label{dirac delta}%
To derive this rigourously, one can make use of the identity
\bea
\lim_{a\to \infty} \, \big({e^{\,a \cosh x+b} - 1}\big)^{-1} &=&
e^{-(a+b)} \sqrt{{2\pi}/{a}} \  \delta(x) \, . \nonumber
\eea
}
Therefore, when integrating the rate over $y\,$, 
like in Eq.~\eq{eq:spect} for the yield, 
we only pick up
a contribution from spacetime coordinates with 
$\eta_s \in [y_{\rm min} , y_{\rm max}]$. 
For such cells, the rate is
\bea
\frac{\dd \Gamma_{\ell \bar \ell}}{\dd M \, \Delta y}  &\simeq&
\frac{\alpha_{\rm em}^2}{9\pi^3 M \, \Delta y} \sqrt{\frac{8\, T}{\pi^5\, \gamma_\perp}}
\,
\int_{k_{\rm min}}^{k_{\rm max}} \frac{\dd k_\perp \, k_\perp}{\sqrt{M_\perp}}
\int_0^{2\pi} \dd\varphi 
\nonu\\
&\times& 
\,
\exp \bigg[
  - \frac{ 
  \omega^\star
  }{T} 
  \bigg]
  \,
  \rho_{\rm V}(\omega^\star, k^\star)
  \, , \label{dGammadMdy approx}
\eea
where $\omega^\star$ and $k^\star$ are given by 
\eq{omega and k boost inv} with $y=\eta_s$.
In the boost invariant setup we are considering, local thermodynamic
variables are functions of only $\bm x_\perp$ and $\tau$
implying that \eq{dGammadMdy approx} is 
flat w.r.t. $\eta_s\,$.

Exploring the rapidity dependence of the yields is particularly 
imperative for low beam energy collisions. 
To emphasize this,
we perform a numerical integration of Eq.~\eq{convolution-bjorken}
under the assumption of boost invariance
and simply set $k_{\rm min} =0$ and $k_{\rm max} = \infty\,$.
Figure~\ref{fig:yields_in_rapidity} shows the
rate which is a (symmetric) function of the rapidity
difference $y-\eta_s$ because of Eq. (\ref{omega and k boost inv}).
For illustration, we set 
 $T=0.25$ GeV and $\muB=0.1$ GeV and $u_\perp=0.3\,$.
This demonstrates that indeed the rate is concentrated 
around $y=\eta_s$ and the width decreases
as a function of the invariant mass.
Given the finite width, 
compared to the limit in footnote~\ref{dirac delta},
it's also crucial to compute dilepton production 
from a (3+1)-dimensional boost-non-invariant system generated in low beam energy collisions. 
This is necessary for making quantitative 
comparisons with experimental measurements 
within a specified rapidity window,
instead of taking the `central value' for $y\,$.

\bigskip
\bibliography{refs}
\end{document}